\documentclass[fleqn,twocolumn,tighten,apj]{aastex63}
\usepackage{amsmath,nccmath,scalerel,multirow,dsfont,balance,gensymb,hyperref,}

\newcommand\HI{H\protect\scaleto{$I$}{1.2ex}}
\received{\today}
\revised{TBD}
\accepted{TBD}

\submitjournal{ApJ}

\shorttitle{WST-LDA MHD Classification}
\shortauthors{Saydjari et al.}

\graphicspath{{./}{figures/}}

\begin{document}

\title{Classification of Magnetohydrodynamic Simulations using Wavelet Scattering Transforms}

\correspondingauthor{Andrew Saydjari}
\email{andrew.saydjari@cfa.harvard.edu}

\author[0000-0002-6561-9002]{Andrew K. Saydjari}
\affiliation{Department of Physics, Harvard University, 17 Oxford St., Cambridge, MA 02138, USA}

\author[0000-0001-8132-8056]{Stephen K. N. Portillo}
\affiliation{DIRAC Institute, Department of Astronomy, University of Washington, 3910 15th Ave. NE, Seattle, WA 98195, USA}

\author[0000-0002-1208-119X]{Zachary Slepian}
\affiliation{Department of Astronomy, University of Florida, 211 Bryant Space Science Center, Gainesville, FL 32611, USA}
\affiliation{Physics Division, Lawrence Berkeley National Laboratory, 1 Cyclotron Rd., Berkeley, CA 947109, USA}

\author[0000-0002-3276-6030]{Sule Kahraman}
\affiliation{Department of Electrical Engineering and Computer Science, Massachusetts Institute of Technology, 77 Massachusetts Ave., Cambridge, MA 02139, USA}

\author[0000-0001-5817-5944]{Blakesley Burkhart}
\affiliation{Department of Physics and Astronomy, Rutgers University, 136 Frelinghuysen Rd., Piscataway, NJ 08854, USA}
\affiliation{Center for Computational Astrophysics, Flatiron Institute, 162 5th Ave., New York, NY 10010, USA}

\author[0000-0003-2808-275X]{Douglas P. Finkbeiner}
\affiliation{Department of Physics, Harvard University, 17 Oxford St., Cambridge, MA 02138, USA}
\affiliation{Harvard-Smithsonian Center for Astrophysics, 60 Garden St., Cambridge, MA 02138, USA}

\begin{abstract}

The complex interplay of magnetohydrodynamics, gravity, and supersonic turbulence in the interstellar medium (ISM) introduces non-Gaussian structure that can complicate comparison between theory and observation. We show that the Wavelet Scattering Transform (WST), in combination with linear discriminant analysis (LDA), is sensitive to non-Gaussian structure in 2D ISM dust maps. WST-LDA classifies magnetohydrodynamic (MHD) turbulence simulations with up to a 97\% true positive rate in our testbed of 8 simulations with varying sonic and Alfv\'{e}nic Mach numbers. We present a side-by-side comparison with two other methods for non-Gaussian characterization, the Reduced Wavelet Scattering Transform (RWST) and the 3-Point Correlation Function (3PCF). We also demonstrate the 3D-WST-LDA and apply it to classification of density fields in position-position-velocity (PPV) space, where density correlations can be studied using velocity coherence as a proxy. WST-LDA is robust to common observational artifacts, such as striping and missing data, while also sensitive enough to extract the net magnetic field direction for sub-Alfv\'{e}nic turbulent density fields. We include a brief analysis of the effect of point spread functions and image pixelization on 2D-WST-LDA applied to density fields, which informs the future goal of applying WST-LDA to 2D or 3D all-sky dust maps to extract hydrodynamic parameters of interest.

\end{abstract}

\section{Introduction} \label{sec:intro}

Interstellar dust and gas indirectly trace ISM turbulence, rendering them a vital observational lever on the processes that shape star formation and galaxy evolution \citep{Elmegreen:2004:ARA&A:,Goodman:2009:Natur:,Padoan:2016:ApJ:,Krumholz:2018:MNRAS:,Burkhart:2015:ApJL:}. Furthermore, dust in the ISM produces a foreground signal that must be removed from extragalactic measurements such as Cosmic Microwave Background (CMB) temperature and polarization anisotropy \citep{PlanckCollaboration:2017:A&A:} and the spectra of galaxies, supernovae, and stars \citep{Cardelli:1989:ApJ:, Corasaniti:2006:MNRAS:}. However, the complex interplay of magnetohydrodynamics (MHD), gravity, and supersonic turbulence introduces non-Gaussian correlations into the density fields, complicating both modeling the dust and its subtraction for cosmology \citep{Kandel:2017:MNRAS:,Kritsuk:2018:PhRvL:}.

A useful statistical starting point for describing (real and mock) ISM density fields is the 2-Point Correlation function (2PCF), or its Fourier-domain analogue, the Fourier power spectrum (PS) \citep{Lazarian:2001:ApJ:,Kowal:2007:ApJ:}. However, the 2PCF cannot describe non-Gaussian processes because the power spectrum ignores phase information, which is important for higher-order correlations \citep{Peek:2019:ApJL:}. Higher-order correlation functions such as the 3-Point Correlation function (3PCF) and its Fourier-domain analogue, the bispectrum, provide an improvement in capturing higher-order correlations \citep{Peebles:2001:ASPC:}. The 3PCF has had significant success in cosmology describing the non-Gaussian distribution of galaxies (\citet{Slepian:2017:MNRAS:a,Slepian:2017:MNRAS:b}, and references therein). The 3PCF  \citep{Portillo:2018:ApJ:} and bispectrum \citep{Burkhart:2009:ApJ:, Burkhart:2016:ApJ:} have also been applied to MHD simulations and show some discriminatory power, but interpretation of these higher order statistics remains difficult.

The wavelet scattering transform (WST) provides another way to go beyond the 2PCF, and has the following convenient properties. It is an approximately translation-invariant image representation, which is linear (Lipschitz continuous) in deformations  \citep{Bruna:2012:arXiv:,Mallat:2011:arXiv:}. This means that if an image is slightly deformed from $i$ to $i'$, the difference in the WST coefficients describing the two images is bounded by the size of the deformation. The success of the WST in image discrimination derives from its encoding non-Gaussianity (higher order correlation functions), which allows the WST to distinguish images with the same Fourier power spectrum (see Figure 5 from \citealt{Bruna:2012:arXiv:}). Further, unlike explicit computation of the higher-order correlations, which are susceptible to large variance arising from outliers,\footnote{An outlier with value $X$ will enter as $X^n$ in the N-Point Correlation Function.} the WST coefficients are computed using non-expansive operators and thus have lower variance.\footnote{A non-expansive operator is an operator $M$ such that $\Vert Mx-My\Vert\leq\Vert x-y\Vert$ where $||x||^2 = \int |x(u)|^2 du$ . In the context of the WST, $M$ is simply the usual modulus for complex numbers which is non-expansive by the triangle inequality. This property ensures stability to additive noise because small changes in the field cannot cause more than the same small changes in the coefficients. Further, an outlier with value $X$ always enters the $n^{\rm th}$-order scattering coefficient proportionally to $X$ (see Section \ref{sec:2DWSTTh}).} The variance of the WST coefficients in response to outliers rapidly decreases as the maximum spatial scale size $J$ increases because outliers are suppressed via spatial averaging. 

The WST performs well on standard classification benchmarks (e.g., the MNIST digit dataset, \citealt{lecun-mnisthandwrittendigit-2010}), even with minimal training data. It is sufficiently robust to temporal instabilities to characterize 1D time series data, like the measured velocity at a point in turbulent gaseous flows \citep{Bruna:2013:arXiv:}. Both 2D and 3D formulations of the WST have been applied to density functional theory (DFT) calculations of molecular structure \citep{Hirn:2016:arXiv:,Eickenberg:2018:JChPh:}. In addition, progress has been made on image reconstruction from the WST coefficients using generative networks \citep{Angles:2018:arXiv:} and gradient descent \citep{Bruna:2018:arXiv:}.

The general class of wavelet techniques have long found applications in astrophysics. Early work analyzed $^{13}$CO velocity fluctuations in star forming regions by wavelet transforms \citep{Gill:1990:ApJL:}. The Wavelet Transform Modulus Maxima (WTMM) approach was used to analyze ISM emission maps, including anisotropies in \HI{} \citep{Khalil:2006:ApJS:} from the Canadian Galactic Plane Survey \citep{Taylor:2003:AJ:} and non-Gaussianity \citep{Robitaille:2014:MNRAS:} in the Herschel infrared Galactic Plane Survey (Hi-GAL) \citep{Molinari:2010:PASP:}. However, the WST is \emph{not} merely a wavelet transform. It is repeated convolution by wavelets and pooling under a non-linear function (the modulus). 

While introduced in the deep learning community for image discrimination and classification, the wavelet scattering transform (WST) has recently made inroads in astrophysics. The WST has found applications in cosmology to state-of-the-art, large-scale structure syntheses and cosmological parameter inference where it outperforms the power spectrum in the context of weak lensing \citep{Allys:2020:arXiv:,Cheng:2020:arXiv:}. Recently, a reduced form of the WST (RWST) was introduced to characterize the ISM with interpretable coefficients. The RWST was subsequently applied to Herschel observations and MHD simulations (in 2D) \citep{Allys:2019:A&A:} as well as polarized dust emission \citep{Regaldo-SaintBlancard:2020:arXiv:}. 

The demonstrated capability of the WST to capture non-Gaussianity, success in image classification problems, and prior applications to astrophysical data motivated us to search for an optimal low-dimensional reduction of the WST for classification of dust density fields to extract relevant hydrodynamic parameters. While convolutional neural nets (CNNs) have been trained to differentiate simulations that are sub/super-Alfv\'{e}nic\footnote{Alfv\'{e}n waves are transverse waves for which magnetic tension provides the restoring force, and the Alfv\'{e}n speed scales as $\sqrt{B}$; one often measures turbulent velocity magnitudes relative the Alfv\'{e}n speed.} \citep{Peek:2019:ApJL:} and entire perpendicular velocity fields from MHD simulations \citep{AsensioRamos:2017:A&A:}, we show that the WST provides ``training-free'' classification when combined with a fast and deterministic dimensional reduction step. 

\section{Methods} \label{sec:Methods}

\subsection{Non-Gaussian Descriptors} \label{sec:NGPTheory}
\subsubsection{3-Point Correlation Function (3PCF)} \label{sec:3PCFTh}
For a given density field $I(\vec x)$, correlations between three points, $I(\vec x_1)I(\vec x_2)I(\vec x_3)$ for all $\vec x_1$, $\vec x_2$, $\vec x_3$, can be used to construct higher-order statistics that capture non-Gaussianity. We generally seek summary statistics that are translation-invariant, so we compute the translation and rotation-averaged product over three points, which yields the 3PCF. The 3PCF then depends only on three coordinates which can be parameterized by $r_{12}$, $r_{13}$, and $\theta$. Here $r_{12}$ is the distance from some base point $\vec x_1$ to $\vec x_2$, $r_{13}$ is the distance from $\vec x_1$ to $\vec x_3$, and $\theta$ is the angle between $\hat r_{12}$ and $\hat r_{13}$. Correlations between $\vec x_1$, $\vec x_2$, $\vec x_3$ include contributions from the 2PCF between pairs of these three points \citep{Groth:1977:ApJ:}. Setting the mean density to zero (i.e. $\int I(\vec x) d\vec x = 0$) removes this dependence of the 3PCF on the 2PCF (see Appendix \ref{sec:3PCF2PCF}). By expanding the angular dependence in terms of multipole moments around $\vec x_1$, this averaged-3PCF can be computed on a regular grid (of $N_{\rm g}$ points) with an FFT-based algorithm in $N_{\rm g} \log(N_{\rm g})$ time \citep{Slepian:2016:MNRAS:}. In 2D, we compute the projection of the 3PCF onto $e^{imx}$ for $m \geq 0$, where the coefficients for $m < 0$ are obtained by complex conjugation. Since the 3PCF is constructed as a triple product of a real field, the 3PCF must be real. However, our projection makes the 3PCF coefficients in general complex-valued. Viewing the Fourier expansion in terms of cosines and sines where all coefficients are real, this is equivalent to saying that we find contributions which are anti-symmetric in $\theta$. This is not possible in 3D because a triangle defined by ($r_{12}$, $r_{13}$, $\theta$) is equivalent to that defined by ($r_{12}$, $r_{13}$, -$\theta$) under rotations and translations. However, in 2D, these triangles can only be identified under reflection.\footnote{We verified that the real and imaginary parts of the 3PCF coefficients are even and odd under reflection as expected.} Since previous works used only the even components of the 2D-3PCF coefficients \citep{Zheng:2004:ApJ:, Chen:2005:ApJ:}, we will provide comparisons using the real part of the coefficients in any case where we make use of the fully complex coefficients. 

We modified an unpublished 2D version of the code described in \citet{Portillo:2018:ApJ:} and increased the speed. The code we used is publicly available (see Section \ref{sec:dataavil}). We characterize a 3PCF computation by the number of radial bins selected for $r_{12}$ and $r_{13}$ and the maximum multipole moment ($L_{\rm max}$) retained in the expansion. A 3PCF with $N_{\rm bins}$ and a maximum multipole moment of $L_{\rm max}$ thus has $(L_{\rm max}+1)N_{\rm bins}(N_{\rm bins}+1)/2$ coefficients; the $L_{\rm max}+1$ comes from multipoles running from $L = 0, ..., L_{\rm max})$ and the factor of two comes from the symmetry $r_{12}$ and $r_{13}$. The 3PCF is averaged over the radii spanned by each bin. For optimal comparison to the WST, we chose eight logarithmic radial bins bounded by $2^j$ for $j = 0, ..., 8$.\footnote{By including $j = 8$, triangles with large radii and large angles can probe pixels that are highly correlated in the presence of periodic boundary conditions.}

\subsubsection{2D Wavelet Scattering Transform (2D-WST)} \label{sec:2DWSTTh}

In 2D, a WST is specified by ($J$, $L$, $M$) where $J$ is the maximum scale, $L$ is the number of angular bins (i.e. angular spacing is $180\degree/L$), and $M$ is the maximum order of the transform. The maximum scale and number of angular bins ($J$, $L$) define the set of Morlet wavelets $\psi_{j,l}$ used in the transform.\footnote{The Wavelet Scattering Transform we consider here is based on a specific instance of wavelet decomposition, which is a more general technique. In fact, a similar wavelet decomposition was previously performed on the MHD simulations we study \citep{Kowal:2010:ApJ:}, though that decomposition was not used as the basis for the WST or as inputs for a classification algorithm.} The parent Morlet wavelet is the product of a plane wave of unit wavenumber with a Gaussian window of width $\sigma$. In one dimension, the wavelet is given by
\begin{ceqn}
\begin{align}\label{eq:WST_2D_wavelet}
\psi(x) & = \alpha(e^{i \frac{3\pi}{4} x}-\beta)e^{-x^2/(2\sigma^2)}
\end{align}
\end{ceqn}
where $\alpha$ and $\beta$ are normalization factors \citep{Ashmead:2010:arXiv:}. We choose $\alpha$ so that the wavelets cover the Fourier domain as uniformly as possible. We choose $\beta$ such that $\psi$ has a null average. 

\begin{figure}[hb!]
\includegraphics[width=\linewidth]{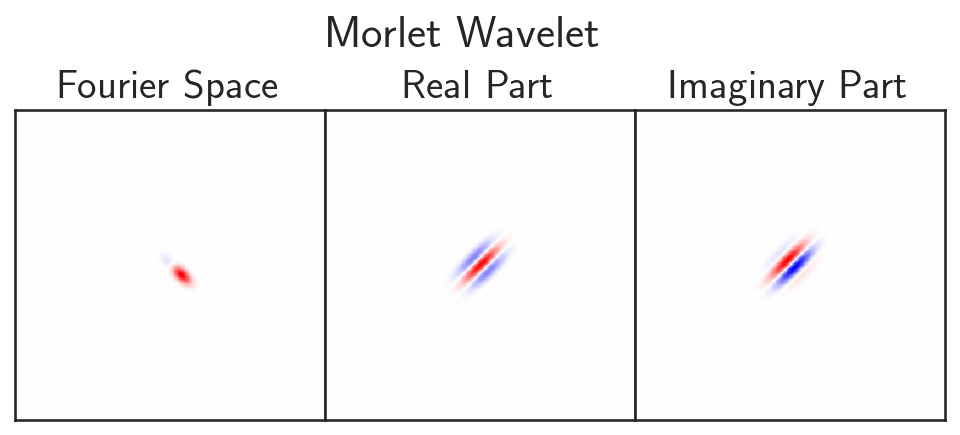}
\caption{Support of an example Morlet wavelet $\psi_{j,l}$ with ($j$,$l$) = (3,5) or ($j$,$\theta$) = (3,$-\pi/4$) for a $256^2$ image in the Fourier domain (\emph{left}). The real (\emph{middle}) and imaginary (\emph{right}) components of the wavelet are shown in the spatial domain. Angles are defined positive clockwise from north. \label{fig:Morlet}}
\end{figure}

In 2D and higher, the inverse variance $1/\sigma^2$ of the Gaussian window becomes an inverse covariance matrix. We choose a diagonal covariance matrix with $\sigma_{yy} = 2\sigma_{xx}$ (where $x$ is the direction of the plane wave) to enhance the directionality of the wavelets. We choose $\sigma_{xx}$ = 0.8 for our 2D WST, which limits the number of oscillations of the plane wave, thereby enhancing filament identification. Each wavelet used in computing the WST coefficients is said to represent an \emph{oriented scale} indexed by $(j,l) \in ([0,...,J-1],[0,...,L-1])$ and is obtained from the parent wavelet by rotation and scaling of the argument (by $2^{j}$) and frequency (by $2^{-j}$).\footnote{While the maximum spatial scale is $2^J$, the wavelet indices refer to the wavevector (in pixels) in the Fourier domain. Thus $j = J-1$ in the Fourier domain is the minimum possible scale for a spatial domain of $2^J$.} While we use $l$ as an index, we denote the corresponding angles as $\theta \in [\pi/2-\pi/L, ..., -(L-1)\pi/2L, -\pi/2]$. The wavelet $\psi_{j,l}$ has a bandwidth of $\sim 2^{-j}$ is centered at a wavevector of magnitude $\sim2^{-j}$ and angle $\theta$ corresponding to $l$ in the Fourier domain (Figure \ref{fig:Morlet}). 

The WST can be represented as a convolutional network with layers labeled $m$ from $0$ to $M$. We will outline the computation of coefficients for each layer for a network with $M=2$. Coefficients for the $m^{\rm th}$ layer are derived from $m$ convolutions of the image field $I(\vec x)$ with Morlet wavelets under the usual modulus for complex numbers. Normalization of the coefficients is relative to the response to a Dirac delta distribution $\delta_{2\rm{D}}(\vec x)$.
\begin{align}
S_0 & = \frac{1}{\mu_0} \int I(\vec x) \;d^2\vec x \qquad\\ \nonumber
\mu_0 & = \int |\delta_{2\rm{D}} \star \mathds{1}_\mathds{R}|(\vec x) \;d^2\vec x\\ \nonumber
S_1(j_1,l_1) & = \frac{1}{\mu_1} \int |I \star \psi_{j_1,l_1}|(\vec x) \;d^2\vec x \\ \nonumber
\mu_1 & = \int |\delta_{2\rm{D}} \star \psi_{j_1,l_1}|(\vec x) \;d^2\vec x\\ \nonumber
S_2(j_1,l_1,j_2,l_2) & = \frac{1}{\mu_2} \int ||I \star \psi_{j_1,l_1}| \star \psi_{j_2,l_2}|(\vec x) \;d^2\vec x \\ \nonumber
\mu_2 & = \int ||\delta_{2\rm{D}} \star \psi_{j_1,l_1}| \star \psi_{j_2,l_2}|(\vec x) \;d^2\vec x
\end{align}
Here $\mathds{1}_\mathds{R}$ is the indicator function, which is $1$ everywhere in the image. For $\mu_0$ and $\mu_1$ the convolution simplifies to the area of the image and the sum of $|\psi_{j_1,l_1}|$ over the image respectively.\footnote{Note that this formal normalization is not implemented in \textsc{Kymatio}.} Note that first-order coefficients depend on one oriented scale $(j_1,l_1)$, while second-order coefficients depend on two oriented scales $(j_1,l_1)$ and $(j_2,l_2)$. Since the first convolution obscures information at scales below $2^{j_1}$ pixels, only coefficients with $j_2 > j_1$ are computed, because then one is only looking at scales $2^{j_2} > 2^{j_1}$. There is one coefficient for $m=0$, $JL$ coefficients for $m=1$, and $J(J-1)L^2/2$ coefficients for $m=2$ for a total of $J(J-1)L^2/2 + JL + 1$. The WST coefficients at the $m^{\rm th}$ layer depend on correlation functions up to order $2^m$ (proof in \citealt{Mallat:2011:arXiv:}). Thus, $M > 1$ is required to capture non-Gaussianity because for $m = 1$, the WST coefficients depend predominantly on the 2PCF. For many applications, the spectral energy falls off rapidly with $m$, and $M=2$ provides sufficient sensitivity to non-Gaussianity with a tractable number of coefficients. 

We normalize the 2D WST coefficients following \citet{Allys:2019:A&A:} and \citet{Bruna:2013:arXiv:} as
\begin{align}
\bar S_0 & = \log_2\left[S_0\right] \nonumber\\
\bar S_1(j_1,l_1) & = \log_2\left[S_1(j_1,l_1)/S_0\right] \\ 
\bar S_2(j_1,l_1,j_2,l_2) & = \log_2\left[S_2(j_1,l_1,j_2,l_2)/S_1(j_1,l_1)\right] \nonumber
\end{align}
This normalization makes the first- and second-order coefficients invariant under multiplication of the field by a constant factor and the second-order coefficients invariant to modification of the spectrum of the field by the action of a class of linear filters \citep{Bruna:2018:arXiv:}.

\newpage
\subsubsection{3D Wavelet Scattering Transform (3D-WST)}
In 3D, the WST coefficients are defined similarly, except that solid harmonic wavelets (denoted $\psi_l^m$) are used instead of Morlet wavelets, with $m \in [-l,...,l-1,l]$.
\begin{ceqn}
\begin{align}\label{eq:WST_3D_wavelet}
\psi_l^m(\vec x) = \frac{1}{(\sqrt{2\pi})^3}e^{-|\vec x|^2/2}|\vec x|^l Y_l^m\left(\frac{\vec x}{|\vec x|}\right)
\end{align}
\end{ceqn}
The $Y_l^m$ are the Laplacian spherical harmonics, with $(j,l) \in ([0,...,J-1,J],[0,...,L-1,L])$. This corresponds to the choice of $\sigma = 1$ for the Gaussian window; the $j$-dependent wavelets are obtained by scaling the argument, as in 2D. We follow the implementation in \textsc{kymatio}\footnote{\textsc{kymatio} \citep{Andreux:2018:arXiv:} is a \textsc{Python} implementation of the WST available at \url{https://www.kymat.io}} \citep{Eickenberg:2018:JChPh:}, averaging over $m$ in computing both the first- and second-order coefficients. For the first-order coefficients this is explicitly
\begin{align}
S_1(j_1,l_1) & = \frac{1}{\mu_1} \int \left(\sum_{m = -l_1}^{l_1}|I \star \psi_{j_1,l_1}^m(\vec x)|^2\right)^{q/2} d^3\vec x
\end{align}
where $q = 1$. For the 3D-WST, $S_2(j_1,j_2,l_1)$ is a function of one angular scale only (we take $l_2 = l_1$). This means the second order coefficients are obtained between scales at the same angles only (instead of the Cartesian product of all angles). We compute the convolution coefficients for $q = 1/2$ and $q = 2$ in addition to $q = 1$, which is the exclusive power used in 2D WST, though we comment only briefly on the $q \neq 1$ cases.\footnote{For $q = 1/2$, small nonzero values are upweighted providing a measurement of sparsity. For $q = 1$, the WST coefficients just scale linearly with density. For $q = 2$, $\left[\rm density\right]^2$ interactions are upweighted.} There are $(L+1)(J+1)$ coefficients for $m = 1$ and $(L+1)J(J+1)/2$ coefficients for $m = 2$ for a total of $(L+1)(J^2+3J+2)/2$ (for each power $q$ computed).

\subsubsection{Reduced Wavelet Scattering Transform (RWST)}
The RWST reduction (which is only defined in 2D) takes advantage of periodicity observed in the WST coefficients to remove the angular dependence (see Appendix \ref{sec:GRFosc}). Explicitly, the RWST coefficients are obtained from the WST coefficients by least-squares fit of the first and second order coefficients. For first order,
\begin{align}\label{eq:RWST1}
\bar S_1(j_1,l_1) = & S_1^{\rm iso}(j_1) + \nonumber\\
& S_1^{\rm aniso}(j_1) \cos\left[\frac{2\pi}{L}(l-l_{\rm a}^{\rm ref})\right]
\end{align}
where $S_1^{\rm iso}(j_1)$, $S_1^{\rm aniso}(j_1)$, and $l_{\rm a}^{\rm ref}(j_1)$ are fit coefficients. For second order, 
\begin{align}\label{eq:RWST2}
\bar S_2(j_1,l_1,j_2,l_2) = & S_2^{\rm iso,1}(j_1,j_2) + \\
& S_2^{\rm iso,2}(j_1,j_2) \cos\left[\frac{2\pi}{L}(l_1-l_2)\right] + \nonumber\\
& S_2^{\rm aniso,1}(j_1,j_2) \cos\left[\frac{2\pi}{L}(l_1-l_{\rm b}^{\rm ref})\right] + \nonumber\\
& S_2^{\rm aniso,2}(j_1,j_2) \cos\left[\frac{2\pi}{L}(l_2-l_{\rm b}^{\rm ref})\right] \nonumber
\end{align}
where $S_2^{\rm iso,1}(j_1,j_2)$, $S_2^{\rm iso,2}(j_1,j_2)$, $S_2^{\rm aniso,1}(j_1,j_2)$, $S_2^{\rm aniso,2}(j_1,j_2)$, and $l_{\rm b}^{\rm ref}(j_1,j_2)$ are fit coefficients. There are $3J$ coefficients for $m=1$, and $5J(J-1)/2$ coefficients for $m=2$, for a total of $J(5J+1)/2$.\footnote{After the completion of this work, we became aware of \textsc{pywst}, a \textsc{Python} implementation of the RWST, available at \url{https://github.com/bregaldo/pywst}. Our own implementation can be found in Section \ref{sec:dataavil}. We compared the two implementations on the same images and obtained similar results, except that \textsc{pywst} includes $S_0$ in the RWST coefficients. Importantly, we also bound the phases to a single interval [0,2$\pi$]. This provided improved convergence but also introduced minor discrepancies between the coefficients obtained by the two methods, most prominently in $S_2^{\rm aniso,1}$, $S_2^{\rm aniso,2}$, and the reference angles themselves.}

\subsection{Dimensional Reduction Algorithms} \label{sec:DimRedTh}

In order to interpret these high-dimensional ($>100$ coefficients), non-Gaussian descriptors, we employ dimensional reduction techniques, including principal component analysis (PCA), linear discriminant analysis (LDA), and quadratic discriminant analysis (QDA).\footnote{Previous attempts to reduce the dimensionality of the bispectrum have been made \citep{Burkhart:2016:ApJ:}, but the application was ``not straightforward.''} PCA is an unsupervised technique in which the $i^{th}$ principal component is orthogonal to the first $i-1$ principal components and in the direction of maximal remaining variance \citep{Jolliffe:2016:RSPTA:}. In contrast, LDA and QDA are supervised techniques, meaning they take class labels (here hydrodynamic parameters) in addition to the high-dimensional descriptor of a data point (here the WST, RWST, or 3PCF coefficients). LDA assumes that each component of the high-dimensional descriptor is normally distributed and that all classes share the same covariance matrix; this latter assumption is known as homoscedasticity \citep{Bandos:2009:ITGRS:}. LDA then finds the hyperplanes that best separate the $K$ classes in the $K-1$ dimensional space containing the class means. The hyperplanes are found by eigenvalue decomposition involving the inter- and intra-class scatter matrices. This is equivalent to evaluating the Gaussian likelihood that each input is in each class, and assigning the input to the class with the greatest likelihood. Note that the LDA components (which maximize the separation between classes) need not be orthogonal. The LDA components are in units of $\sigma_{ii}$, the on-diagonal components of the shared covariance matrix. QDA is similar to LDA except that it drops the assumption that the covariance matrix must be the same for all classes and thus finds conic sections that best separate the classes instead of hyperplanes. The boundary between two neighboring classes occurs where the Gaussian likelihoods of each class are equal, which is in general a quadratic function (as in QDA) but simplifies to a linear one when the class covariances are assumed equal (as in LDA). 


\section{Datasets}

In this section, we introduce the observational datasets and MHD simulations used to verify the sensitivity of our method to non-Gaussianity and test its capabilities for classification of turbulence.

\begin{figure}[hb!]
\includegraphics[width=\linewidth]{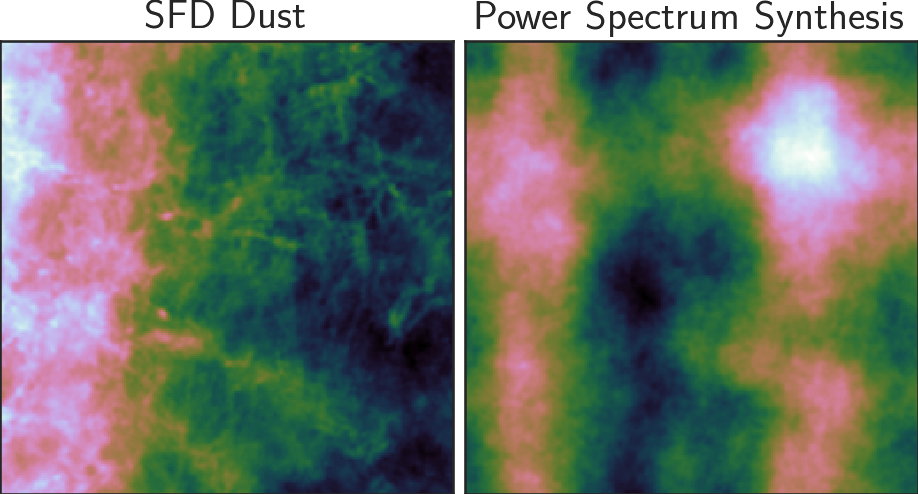}
\caption{\emph{Left}: Representative $256^2$ image of the 2D SFD dust map (log color scale). \emph{Right}: Same as left, but with random phase in the Fourier domain (linear color scale). High intensity (light colors) corresponds to large amounts of extinction and hence high dust density while low intensity (dark colors) corresponds to low dust density. A log color scale is used only for the SFD image to emphasize filament-like structures present in SFD but not present in the power spectrum synthesis. The relevance of phase coherence for non-Gaussian filamentary structure is apparent by comparison. Units are obscured due to the [min,max] = [0,1] normalization adopted.  \label{fig:SFD}}
\end{figure}

\subsection{SFD Dust Map} \label{sec:SFD}

Interstellar dust is an indirect tracer of interstellar turbulence and provides an observational point of comparison to the MHD simulations discussed below. Dust grains are heated to $\sim 20$K by ambient starlight and produce thermal emission in the far-infrared \citep{Low:1984:ApJL:}. We use the SFD dust map \citep{Schlegel:1998:ApJ:}, which is based on this thermal emission.  The map derives dust morphology from $100\mu$m data from the Infrared Astronomical Satellite (IRAS) Sky Survey Atlas \citep{Neugebauer:1984:ApJL:,Wheelock:1994:STIN:}, corrected for temperature variation with data from the Diffuse InfraRed Background Experiment (DIRBE) on the Cosmic Background Explorer (COBE) \citep{Mather:1982:OptEn:,Boggess:1992:ApJ:}. 

To obtain a representative sample of 2D dust images, we reproject the SFD dust map with bilinear interpolation to a TAN (gnomonic) projection centered on random locations on the celestial sphere.  Each $256^2$ pixel image has a random position angle, a pixel scale of 3.0 arcmin, and a point spread function (PSF) full width at half max (FWHM) of 6.1 arcmin.  The SFD bitmask flags regions with missing IRAS data (bit 128), pixels in the Magellanic clouds or M31 (bit 64), and regions with no point-source subtraction (mostly at low Galactic latitude, $|b| < 5^{\circ}$; bit 32).  An image is discarded if \emph{any} pixel has \emph{any} of these mask bits set, leaving a sample of 10,222 images (Figure \ref{fig:SFD}). By rejecting images near the Galactic plane, we focus on morphology of relatively nearby dust (of order a few 100 pc to a few kpc) where each image is dominated by dust at one or a few distances. The statistics of the dust distribution near $b=0$ might be substantially different, owing to the large number of clouds along each line of sight, and the steep dependence on latitude. The images are normalized to have [min,max] = [0,1] for input to the WST and zero mean with unit variance for input to the 3PCF, consistent with other images used in the paper.

\subsection{Magnetohydrodynamic Simulations} \label{sec:MHDSims}

We use MHD simulations of turbulent gas flows published previously \citep{Portillo:2018:ApJ:,Cho:2003:MNRAS:,Burkhart:2009:ApJ:,Bialy:2020:ApJL:,CATSkeyunpub} and review only key properties. The simulations solve the ideal MHD equations constraining the divergence to zero ($\nabla \cdot \vec B = 0$), assuming an isothermal equation of state, and periodic boundary conditions. No gravity is included and the simulations are ``scale-free.'' The simulation box is $256^3$ voxels and is driven isotropically by random, divergence-free forcing with wave number $k$ $\approx$ 2.5 (i.e. 1/2.5 the box size). No dissipation is assumed other than numerical dissipation from the finite resolution of the simulation. The magnetic field is initialized uniformly in the $x$-direction.

\begin{figure}[ht!]
\includegraphics[width=\linewidth]{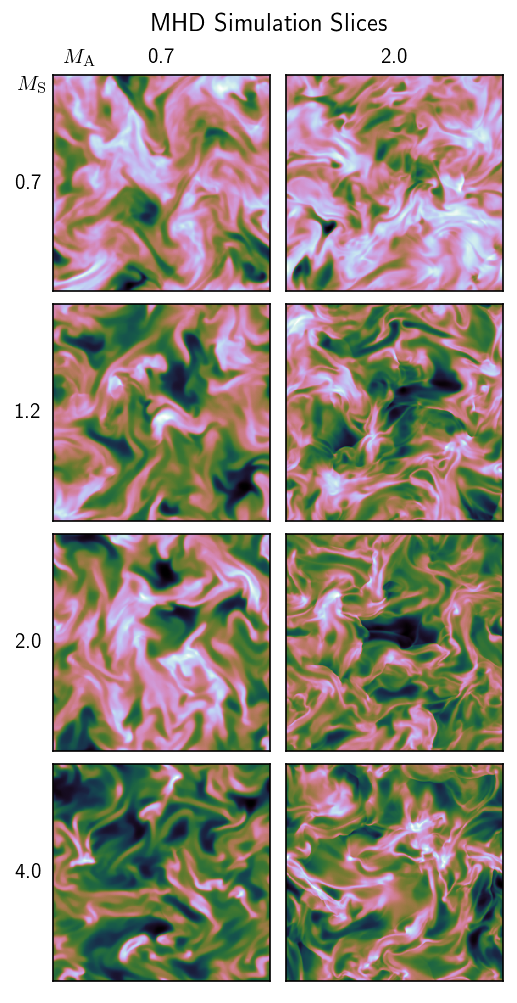}
\caption{Representative $256^2$ slices of the density field for MHD simulations at various Alfv\'{e}nic ($M_{\rm A}$) and sonic Mach numbers ($M_{\rm S}$). The image axes are spatial coordinates at a fixed value for the third spatial coordinate of the simulation box. Density units are obscured by the [min,max] = [0,1] normalization adopted. High intensity (light colors) corresponds to high density and low intensity (dark colors) corresponds to low density. A log color scale is used to emphasize density variance across spatial scales. Plotted on the same scale, larger sonic Mach numbers show larger density fluctuations, but this information is rescaled by normalization prior to computing the relevant statistic. \label{fig:MHD}}
\end{figure}

Each simulation is characterized by its Alfv\'{e}nic Mach number ($M_{\rm A}$) and sonic Mach number ($M_{\rm S}$). The Alfv\'{e}nic Mach number is defined as $M_{\rm A} \equiv |v|/\langle v_A \rangle$ where $v$ is the flow velocity and $\langle v_A \rangle$ the Alfv\'{e}n (magnetic wave) speed averaged over the simulation box. The sonic Mach number is defined as $M_{\rm S} \equiv |v|/c_s$ where $c_s$ is the isothermal speed of sound. We use $M_{\rm A}$ = 0.7, 2.0 and $M_{\rm S}$ = 0.7, 1.0, 2.0, 4.0 to include sub-Alfv\'{e}nic ($M_{\rm A} = 0.7$) and super-Alfv\'{e}nic ($M_{\rm A} = 2.0$) cases as well as subsonic ($M_{\rm S} = 0.7$), transonic ($M_{\rm S} = 1.0$), and supersonic ($M_{\rm S} = 2.0, 4.0$) cases (Figure \ref{fig:MHD}). The images are normalized to have [min,max] = [0,1] for compatibility with the log normalization of the WST, but we have confirmed that our WST classification result is reproduced if one uses zero mean and unit variance normalization. The images are normalized to have zero mean and unit variance when computing the 3PCF in order to exclude any contributions from the 2PCF (see Appendix \ref{sec:3PCF2PCF}).

\section{Results and Discussion} \label{sec:Results}

We begin by comparing the performance of the WST, RWST, and 3PCF in capturing non-Gaussianity. Then, we combine these statistics with LDA to classify turbulent MHD simulations and investigate how robust WST-LDA is to modifications of the sample images.

\subsection{Sensitivity to Non-Gaussianity} \label{sec:NGPResult}

We assess the performance of the WST, its reduction the RWST, and the 3PCF as high-dimensional descriptors that characterize non-Gaussianity in 2D images. Each of these descriptors is then used in conjunction with two dimension-reduction techniques, PCA and LDA. PCA and LDA are used to probe the information contained in the high-dimensional parameter spaces corresponding to the WST, RWST, and 3PCF coefficients. We begin by confirming that these descriptors are sensitive to non-Gaussianity by applying them to 2D dust map images. We compute the WST coefficients ($J = 8$, $L = 8$, $M = 2$) on 10,222 SFD images (``SFD''),\footnote{The density of dust in the ISM is sometimes assumed to be a log-normal distribution; thus we also verified these results on $\log\left[{\rm density}\right]$} 2,000 images where each pixel is a uniformly drawn pseudo-random number [0,1) (``Rand''), and those same random images convolved with a $\sigma = 4$ pixel uniform Gaussian (``Smooth''). This latter set imposes nearest-neighbor correlations, which are present in empirical images because of the continuity of real density fields and instrumental point-spread functions. To further prove that the WST captures non-Gaussianity, we compute the WST coefficients for images that have Fourier power spectra matching the SFD images, but small higher-order correlations (``NHC''). Precisely, for each image in the SFD set, we apodize the image (multiply by a generalized Gaussian,\footnote{A generalized Gaussian has the form $f(x) = e^{{-(|x|/\sigma)}^{2p}/2}$ with a FWHM of $2[2\log(2)]^{1/2p}\sigma$. Note that for $p = 0.5$ this is the Laplace distribution and for $p = 1$ this is the Gaussian distribution.} $p = 6$, $\sigma = 100$), compute the Fourier transform (FT), randomize the phase (with a unique pseudo-random seed), and inverse FT.\footnote{We confirmed that the results obtained for ``SFD'' do not change for apodized ``SFD'' images. This implies that the results for ``NHC'' stem from phase randomization and are not just an effect of apodization.} The RWST and 3PCF ($N_{\rm bins}$, $L$) = (8, 9) coefficients are also computed on the same four datasets for comparison.

\begin{figure}[!hb]
\includegraphics[width=\linewidth]{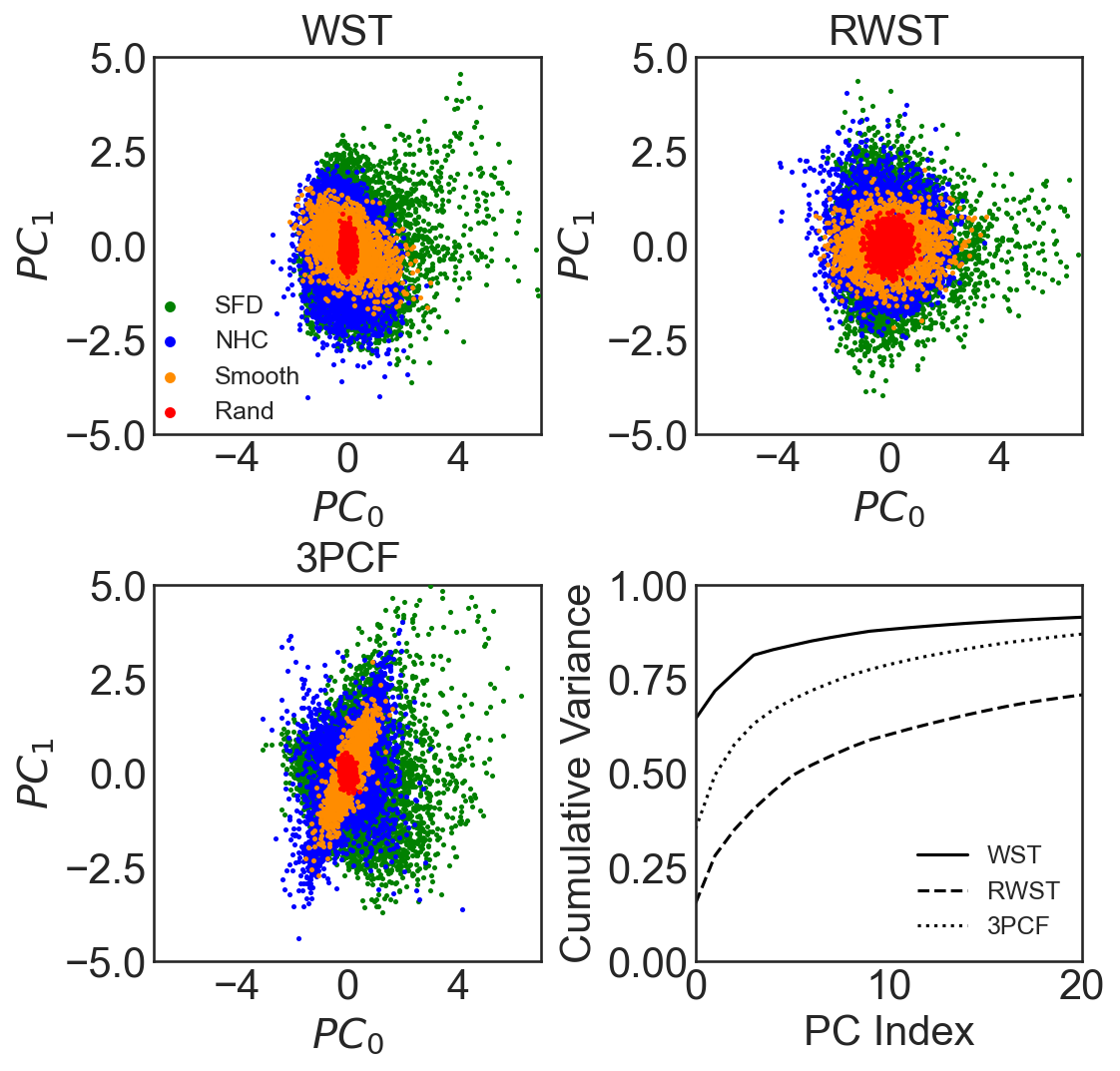}
\caption{Individual image representations in the first two SFD principal components (PCs) are shown for ``SFD'' (green), ``NHC'' (blue), ``Smooth'' (yellow), and ``Rand'' (red). We define these PCs for the WST (\emph{top left}), RWST (\emph{top right}), and 3PCF (\emph{bottom left}) coefficient representations for each image. PC axes are reported in units of the standard deviation of that PC. The \emph{bottom right} panel shows the cumulative variance of the ``SFD'' set explained by the first $N$ principal components (PC) for each statistic. We see the cumulative variance saturate around 5-10 PCs, arguing that not many principal components are required to capture the majority of the variance. An interactive version of this figure is available at \url{https://faun.rc.fas.harvard.edu/saydjari/RWST_2020/SFD_PCAs.html} showing partial separation of the four classes when inputs to PCA are not scaled to have zero mean and unit variance.\label{fig:GRF_PCA}}
\end{figure}

To investigate the variance of these coefficients on prototypical 2D dust, we perform PCA\footnote{We scale the coefficient inputs to PCA to have zero mean and unit variance to account for any differences in magnitude between the coefficients.} on the coefficients from the SFD set and then project the results from the other three control image sets (``Rand'', ``Smooth'', ``NHC'') onto those principal components. Each point in Figure \ref{fig:GRF_PCA} corresponds to an image from one of the data sets above. 

In WST-PCA space, SFD presents as an ellipsoid, with a population of outliers extending along the first PC axis in the positive direction. The ``NHC,'' ``Smooth,'' and ``Rand'' datasets present as progressively narrower ellipsoids that overlap the main mass of ``SFD.'' While slight, the different orientation of these ellipsoids may encode more detailed information differentiating the datasets. We observe that, to first order, the diameter of the ellipsoids decreases with the progressively decreasing pixel-pixel correlations in the ``SFD''$>$``NHC''$>$``Smooth''$>$``Rand'' series. This suggests that the variance in the WST coefficients on ``SFD'' images primarily arises from higher-order correlations as desired. As shown in the cumulative variance plot, most of the variance of ``SFD'' is described by the first few PCA components, suggesting WST-PCA admits a low-dimensional representation of SFD dust. By cross-referencing the Galactic coordinates of the images associated with the outliers in the ``SFD'' dataset using \textsc{glue},\footnote{\textsc{glue} \citep{Beaumont:2015:ASPC:,Robitaille:2017:zndo:} is a \textsc{Python} package for data visualization and linking available at \url{https://glueviz.org/.}} we associated most of the outliers with images of the Orion nebula. The formation of massive stars in Orion modifies the dust distribution, likely causing outliers in the WST-PCA space. This was true for the outliers in the RWST-PCA and 3PCF-PCA bases as well. The sensitivity of WST-PCA to outliers may suggest an application to outlier detection and flagging in large survey pipelines.

When the input coefficients to PCA are not scaled, the mean and variance of the WST coefficients are sufficient to partially separate all four datasets in WST-PCA space (see the interactive version of Figure \ref{fig:GRF_PCA}). This is in contrast to what is expected for a simple N-Point Correlation Function, which should have expectation value zero for $N > 2$ on ``NHC'', ``Smooth'', and ``Rand.'' While exclusion of the $S_0$ coefficient does not visibly inhibit the class separation, we cannot rule out that information about the mean of the density distribution does enter the higher-order coefficients because of the [min,max]=[0,1] normalization we chose. Thus, we believe a more careful study of the distribution and expectation values for the WST coefficients is called for in future work.

The distributions of the four datasets under RWST-PCA are similar to the distribution under WST-PCA except that the ellipsoidal distributions have lower eccentricity. The unscaled RWST-PCA basis also partially separated the four classes and further improved the separation of ``Rand'' to have no overlap with the other three classes. The cumulative variance rises much more slowly as a function of the number of principal components, suggesting that the RWST does not admit as low a dimensional representation and that the variance in ``SFD'' under the RWST is similar for many directions.

We select the real part of the 3PCF coefficients to analyze by PCA. The distribution of the four datasets under 3PCF-PCA is similar to under WST-PCA except that ``Smooth'' and ``Rand'' have higher and lower eccentricities respectively. The unscaled 3PCF-PCA basis separates ``SFD'' from the other three datasets with little overlap. However, the expectation value of the 3PCF for ``NHC'',``Smooth'', and ``Rand'' must be zero. Thus, the centroids for all three datasets are coincident. Despite overlapping centroids, the major axes of ``NHC'' and ``Smooth'' are distinct, allowing partial separation. The distribution of ``Rand'' is so tight and circular that it overlaps entirely with ``NHC'' and ``Smooth.'' The cumulative variance of the 3PCF is between that of the WST and RWST, suggesting the 3PCF admits an intermediate-dimensional representation of SFD dust. We repeated this analysis taking the modulus of the 3PCF coefficients instead. While the exact shape of the distribution for each dataset changed and the centroid of ``NHC'',``Smooth'', and ``Rand'' shifted, the separation was not strongly affected.\footnote{We also ran parallel analyses with only the imaginary component, and a vector doubled in length, including both the real and imaginary components. The imaginary component strongly resembled the real component, except the cumulative explained variance was lower, below that explained by the RWST. Analysis with the doubled vector resembles that of the real component almost exactly.}

The lack of complete separation in the PC basis does not mean these descriptors do not contain the information necessary to distinguish the four classes. PCA is an unsupervised technique and thus cannot optimize for the separation of classes as it has no knowledge of them. To determine if the computed coefficients can distinguish the four classes, we turn to a supervised learning technique, LDA. We split the datasets into test/train sets (80\%/20\%) and compute the confusion matrix for the test set between the true labels and the assignment predicted by the LDA model, which was trained on the first set.\footnote{Just as for PCA, we standard-scale the coefficients prior to LDA.} In the confusion matrix, we show the actual number of images with a true class label that were predicted to have a certain class label. The grayscale coloring of the matrix indicates the true positive percentage relative to the number of test images with the corresponding true label (the row sum). To quantify the predictive power of LDA on this multi-class problem, we report the precision score, which is the total fraction of correctly classified images (true positive percentage).

\begin{figure}[!ht]
\includegraphics[width=\linewidth]{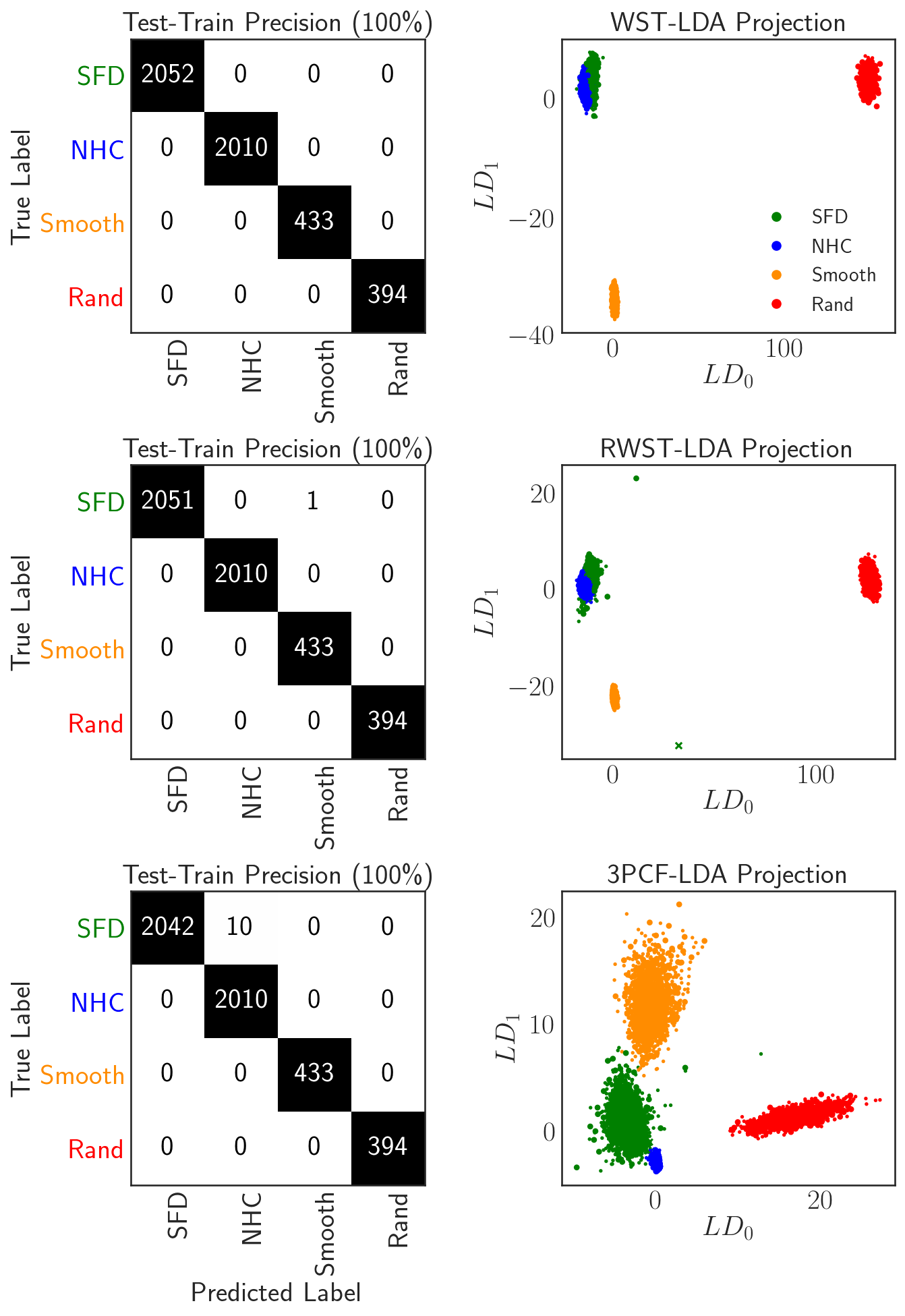}
\caption{Confusion matrix between ``SFD'', ``NHC'', ``Smooth'', and ``Rand'' when applying WST-LDA (\emph{top left}), RWST-LDA (\emph{middle left}), and 3PCF-LDA (\emph{bottom left}). Matrix entries are the number of test images with the corresponding true-predicted label pairing. The grayscale color indicates the true postitive percentage relative to the number of test images with the corresponding true label. Test images are shown in the first two components of LDA space for the WST (\emph{top right}), RWST (\emph{middle right}), and 3PCF (\emph{bottom right}). Circle markers represent images that were correctly assigned and cross markers represent those incorrectly assigned. In all cases, colors correspond to true labels. An interactive version of this figure is available at \url{https://faun.rc.fas.harvard.edu/saydjari/RWST_2020/SFD_LDA.html} showing the images in LDA space and allowing isolation of the train, correctly labeled test images, and incorrectly labeled test images. \label{fig:WST_LDA_GRF}}
\end{figure}

Figure \ref{fig:WST_LDA_GRF} illustrates that the WST contains enough information to completely distinguish the four classes and thus at least partially characterizes the non-Gaussianity of SFD dust. The RWST performs similarly well except for the presence a few outliers in the RWST-LDA space that are not associated with any of the clusters. We speculate that these outliers result from poor convergence in the least-squares fits used to extract the RWST coefficients. While not shown, we also analyzed an RWST model where the frequency in Equations \ref{eq:RWST1} and \ref{eq:RWST2} were fit in addition to the usual RWST parameters. This unfortunately created a large degeneracy between the reference angle and frequency, which caused fitting instability and massive variability that was common to all datasets (not discriminatory). These outliers may arise when the cosinusoidal dependence is so small in amplitude that the angle is undefined, or when the cosinusoidal frequency differs from the model. These points demonstrate that more work remains to regularize the RWST fits.

For improved classification, we take the modulus of the 3PCF coefficients as inputs to LDA. Under these conditions, 3PCF-LDA almost completely distinguishes the four classes, with only a few images of ``SFD'' confused for ``NHC.'' Viewed in the interactive version of Figure \ref{fig:WST_LDA_GRF}, these confused images are associated with the ``SFD'' cluster, suggesting that the misclassification derives from the strictly linear boundaries to which LDA is limited. In addition, the assumption of equal class covariances (homoscedasticity) is strongly violated for the 3PCF because the variance is strongly class dependent (Figure \ref{fig:WST_LDA_GRF}, \emph{bottom right}). To attempt to further improve classification with 3PCF, we use QDA, which relaxes the homoscedasticity condition and unsurprisingly found no misclassified images. When attempting classification with only the real part of the 3PCF coefficients, there was a large confusion between ``NHC,'' ``Smooth,'' and ``Rand,''\footnote{Test-train precision for real part of the 3PCF coefficients was 84\%} which are forced to have exactly overlapping centroids. This condition is lifted upon taking the modulus since the expectation value of the modulus of the 3PCF coefficients need not be zero on images drawn from Gaussian distributions. 

In summary, all three non-Gaussian descriptors contain enough information to distinguish ``SFD,'' ``NHC,'' ``Smooth,'' and ``Rand,'' but with the WST the classes can be distinguished using a simple linear classifier (in contrast to classification with the 3PCF) and without outliers (in contrast to classification with the RWST).

\begin{figure*}[!ht] \centering
\includegraphics[width=0.85\textwidth,height=0.85\textheight,keepaspectratio]{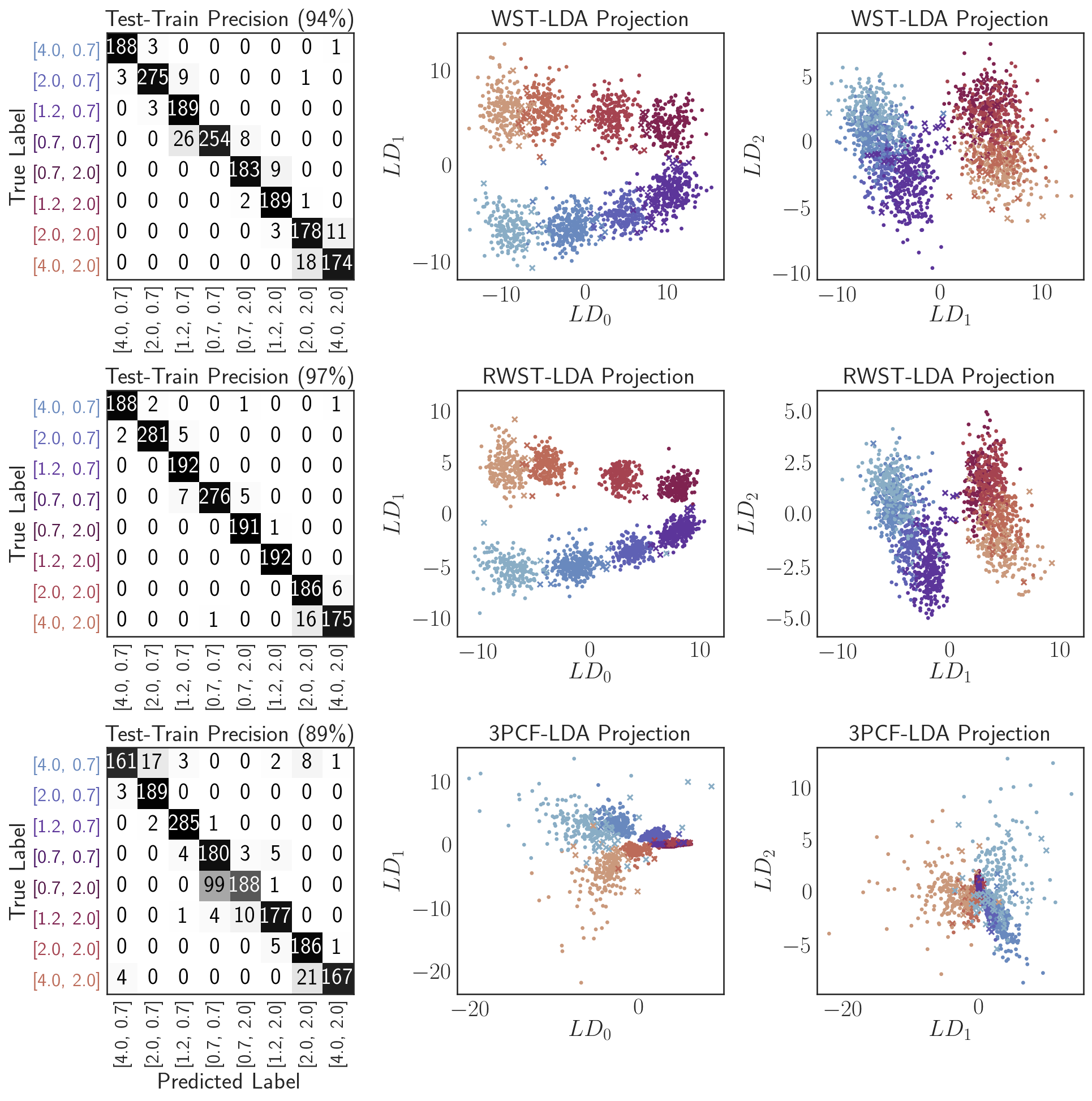}
\caption{
Confusion matrix between MHD classes labeled by [sonic, Alfv\'{e}nic] Mach numbers when applying WST-LDA (\emph{top left}), RWST-LDA (\emph{middle left}), and 3PCF-LDA (\emph{bottom left}). Matrix entries are the number of test images with the corresponding true-predicted label pairing. The grayscale color indicates the true positive percentage relative to the number of test images with the corresponding true label. In all cases, colors correspond to true labels, as indicated on the vertical axis in the left panels. Test images are shown in the first two components of LDA space for the WST (\emph{top right}), RWST (\emph{middle right}), and 3PCF (\emph{bottom right}). Circle markers represent images that were correctly assigned and cross markers represent those incorrectly assigned.  An interactive version of this figure is available at \url{https://faun.rc.fas.harvard.edu/saydjari/RWST_2020/MHD_LDA.html} showing the images in LDA space and allowing isolation of the train, correctly labeled test images, and incorrectly labeled test images. Two outliers for the RWST panels (\emph{middle right}) are beyond the axis limits. The WST, RWST, and 3PCF in combination with LDA all classify the MHD simulations with high precision, though we prefer the linear spacing of classes in the (R)WST-LDA basis. \label{fig:WST_LDA_MHD}}
\end{figure*}

\subsection{Classification of MHD Simulations} \label{sec:Classification}

Since WST, RWST, and 3PCF distinguish Gaussian from non-Gaussian fields, we seek to use them to classify different physical processes that can contribute to non-Gaussianity. One application with implications for the ISM is inferring characteristic MHD parameters, such as the sonic and Alfv\'{e}nic Mach number. We demonstrate the ability of WST-LDA to do this on idealized MHD simulations (Figure \ref{fig:MHD}) as a proof of concept. For each of the eight conditions (sonic/Alfv\'{e}nic Mach number pairs) we have the full density cube ($256^3$ pixels) at nine timesteps. Because the separation between these timesteps is longer than the eddy turnover time, each can be treated as an independent sample for the given condition. For each cube, we take 2D slices ($256^2$ pixels) along each axis (x, y, z) spaced every eight pixels to reduce the sampling redundancy.\footnote{The correlation between slices (which are 1 pixel thick) does not fall to zero until a spacing of $\sim$50 pixels, so sampling every 8 slices does introduce correlations for samples within the same timestep.} Thus, we computed the WST coefficients for 6,912 ($8\times9\times32\times3$) images, or 864 images per class. To ensure the test set is not correlated with the training set, all images in the test set are selected from two independent timesteps withheld from the training set. Since each timestep is an independent sample, the test and train sets are thus independent.

We achieve a precision of 94\% for WST-LDA ($J = 8$, $ L= 8$, $M = 2$; $N_{\rm coeff} = 1,857$), 97\% for RWST-LDA ($N_{\rm coeff} = 164$), and 89\% for 3PCF-LDA ($N_{\rm bins} = 8$, $L = 9$; $N_{\rm coeff} = 360$) (Figure \ref{fig:WST_LDA_MHD}). The precision for WST-LDA is comparable to that obtained previously for differentiating sub/super-Alfv\'{e}nic simulations after training a neural net \citep{Peek:2019:ApJL:}, but notably required no training except for a deterministic LDA step. As shown in the interactive version of Figure \ref{fig:WST_LDA_MHD}, the first three LDA components separate each of the classes. There is a small amount of cross talk between adjacent classes due to scatter of the WST coefficients of images in the same class. 

Since these MHD parameters are actually continuous variables, the variance of each class can be viewed as an indicator of the uncertainty of an inference of these MHD parameters from a single image. To obtain a rough estimate of this uncertainty, we assume $M_{\rm S}$ varies linearly as a function of separation between clusters of the same $M_{\rm A}$ value. Then, we find that $M_{\rm S}$ only varies $\sim10\%$ within $1\sigma$ of the centroid of each cluster. Thus, for these simulations, we can approximately infer $M_{\rm S} \pm 10\%$. Note that the sonic Mach numbers studied here are lower than the average estimated for molecular clouds in the ISM from CO linewidths ($\sim10-20$, \citet{Kainulainen:2013:A&A:,Kainulainen:2017:A&A:}), but comparable to measurements for 21-cm cold neutral medium (CNM) \citep{Heiles:2003:ApJS:,Burkhart:2010:ApJ:}. However, whether this approximately linear mapping of $M_{\rm S}$ to LDA space holds at larger $M_{\rm S}$ remains an open question. 

In this WST-LDA space, the classes of images with different Alfv\'{e}nic Mach number form a helical structure (Figure \ref{fig:MHD}, interactive version). Along the vertical axis of the helix, the sonic Mach number varies monotonically while the separation in the plane perpendicular to this axis determines the Alfv\'{e}nic Mach number. We can also see this behavior in the 2D projections, where $LD_0$ is the helical axis (Figure \ref{fig:WST_LDA_MHD}). There is little structure as a function of Mach numbers in the $4^{\rm th}$ and higher LDA components, suggesting the classification occurs primarily in a 3D subspace of the WST-LDA basis.

The increased precision for RWST-LDA compared to WST-LDA appears to derive from reduced scatter around the centroids of each cluster, though the centroids are separated by the same distance. Since LDA selects the optimal linear combination of the WST coefficients, the success of RWST is not simply in selecting an optimal subset of the coefficients containing the most information; the RWST is not simply performing dimensional reduction. The nonlinear combination of WST coefficients used by the RWST preferentially selects information relevant to differentiating the MHD classes.

For improved classification, we take the modulus of the 3PCF coefficients as inputs to LDA and find partial separation of the classes. We find the variance is strongly class-dependent, with higher sonic Mach numbers having larger variance and lower sonic Mach numbers having smaller variance. This suggests that the variance of the 3PCF is sensitive to sharp density fluctuations. Further, the spacing of the centroids is highly nonlinear. It is the small spacing between these centroids for $M_{\rm S}=0.7$ that is the source of most of the confused images. This small spacing suggests an important influence from $M_{\rm A}$ enters in the $4^{\rm th}$-order moments that are present in the WST, but not in the 3PCF. As aforementioned, for cold gas in the ISM $M_{\rm S}$ is likely $>0.7$; thus this lack of discriminatory power for low $M_{\rm S}$ may not be a problem for some applications. The fact that gas near young hot stars has lower $M_{\rm S}$ suggests it may be more difficult to apply WST-LDA to active star-forming regions \citep{Gallegos-Garcia:2020:ApJL:}.

Since dust maps such as SFD project a 3D volume along the line of sight down to 2D, we investigate how WST-LDA, RWST-LDA, and 3PCF-LDA perform on projected images by repeating our sampling procedure on the cumulative sum of density along a given coordinate axis. This is equivalent to sampling the column density along the line of sight in the optically thin limit. We computed the WST coefficients for the same number of images as before, 6,912. We find a precision of 54\% for WST-LDA, 81\% for RWST-LDA, and 51\% for 3PCF-LDA (see Appendix \ref{sec:cumSum}, Figure \ref{fig:WST_LDA_MHD_cumsum}). The improvement provided by the nonlinear transforms in the RWST becomes even more clear in the context of cumulative sums, and we anticipate that this will be instrumental to applications in which there are multiple clouds along the line of sight.

Overall, the WST and RWST seem most suited to the continuous inference problem of estimating Mach numbers from measured density fields since $M_{\rm S}$ maps approximately linearly into the LDA classification space, in contrast to the highly non-linear mapping for 3PCF-LDA. While the nonlinear transformations of WST coefficients made by the RWST improve the classification as well as render the classification more robust to cumulative sums, future work should seek an optimal nonlinear transformation of WST coefficients that does not also create outliers. Since we do not yet have such an optimal transformation, we focus here on characterizing the performance of WST-LDA on several test cases. 

\begin{figure}[hb!]
\includegraphics[width=\linewidth]{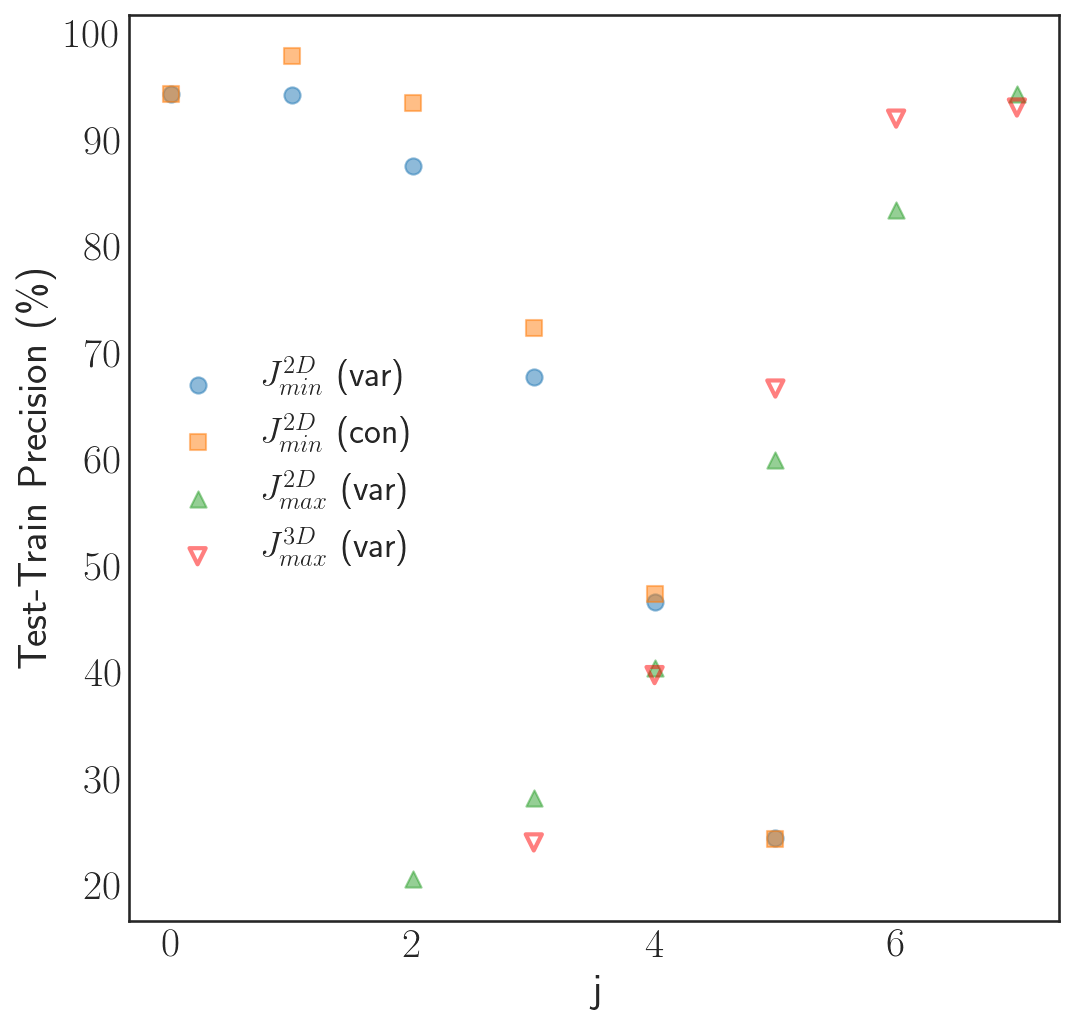}
\caption{Scale-dependent classification precision for WST-LDA on MHD simulations. ``$J_{\rm min}^{2D}$ (var)'' (blue, circle) was obtained using a smaller-$J$ WST after resampling images. ``$J_{\rm max}^{2D}$ (var)'' (green, triangle) was obtained by computing the WST on a $(2^J)^2$ subset of the original images. ``$J_{\rm min}^{2D}$ (con)'' (orange, square) was obtained by computing the WST for $J=8$ on images which were downsampled to $256^2$ after repeated tiling of the image. $J_{\rm min}^{3D}$ (variable) (red, inverted triangle) was obtained by computing the WST on a $(2^J)^3$ subset of the original volumes. Because $J_{\rm min}$ ``constant'' is always within $6\%$ of $J_{\rm min}$ ``variable'', we conclude that the trend observed in $J_{\rm min}$ ``variable'' is predominantly the result of the scales excluded and not the number of coefficients. \label{fig:dim}}
\end{figure}

\subsection{2D/3D Scale Dependence of WST-LDA} \label{sec:Scaling}

It is important to establish how many coefficients the WST needs to characterize a non-Gaussian process. However, this question is intertwined with that of the scale at which the process of interest is occurring and how many different scales the WST needs to characterize a non-Gaussian process. We study this trade-off in our MHD test case by computing the scattering coefficients that include only a subset of the scales in the problem. First we compute the WST coefficients for $L = 8$, $M = 2$, and $J = 3, 4, 5, 6, 7$, and $8$ to investigate the effect of the maximum scale ($J_{\rm max}$).\footnote{Recall the 2D WST convention that $j$ runs from 0 to $J-1$.} We can follow a similar process to find $J_{\rm min}$ by subsampling the $256^2$ image down to a $(2^{8-J_{\rm min}})^2$ image (i.e. binning down by a factor of $2^{J_{\rm min}}$) and computing the WST with $L = 8$, $M = 2$, and $J = 8, 7, 6, 5, 4$, and $3$. Both results are shown in Figure \ref{fig:dim}. Note that the $J_{\rm min}$ (var) and $J_{\rm max}$ (var) curves cross at $j = 4$. The $j = 0$ (smallest spatial) scale likely contains little information if the image is well sampled. Above the $j = 0$ scale, the linearity of each of the datasets suggests that there is approximately the same information content in each scale.

\begin{figure}[hb!]
\includegraphics[width=\linewidth]{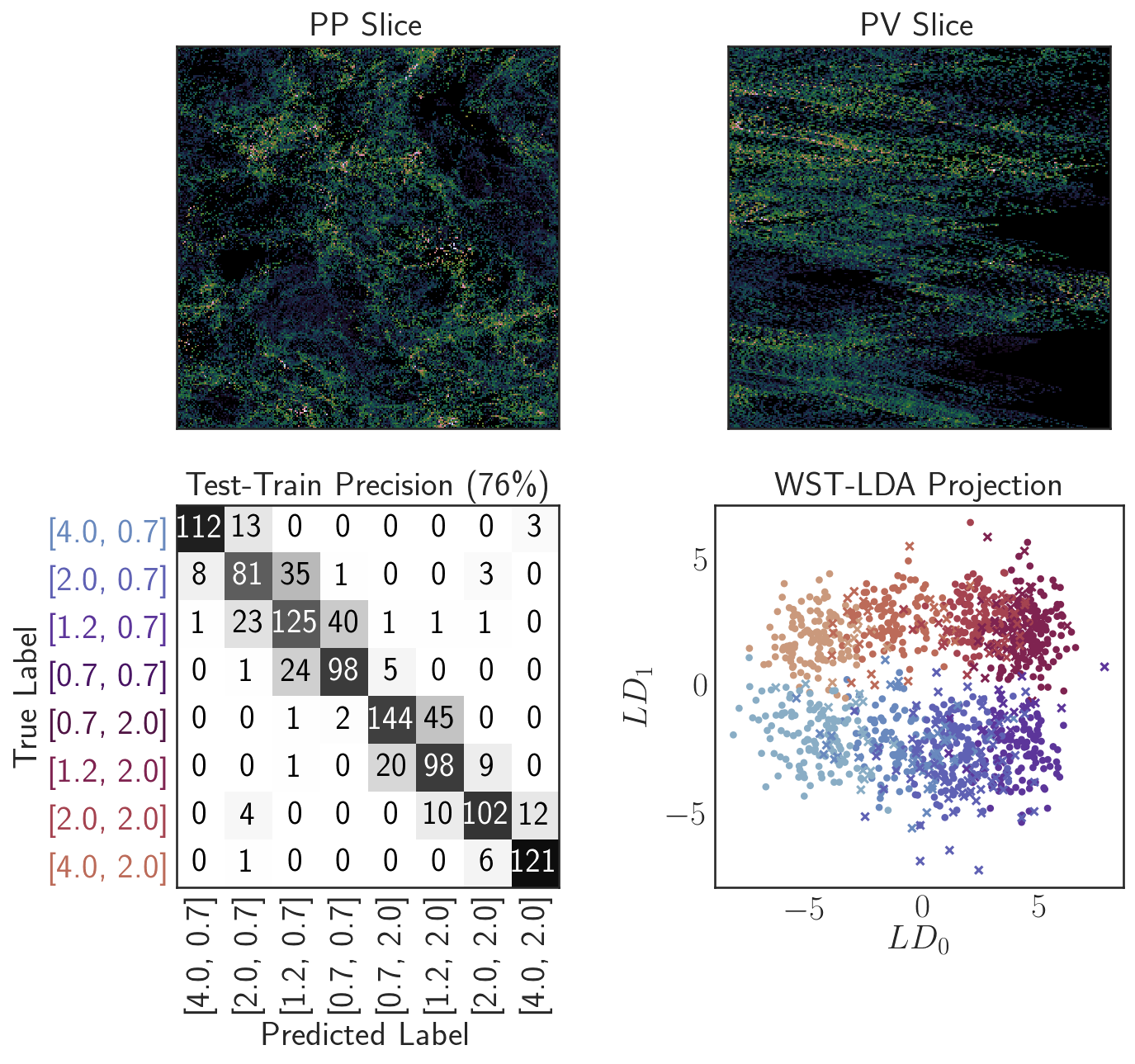}
\caption{Representative position-position (PP) slice for fixed velocity (\emph{top left}) and representative position-velocity (PV) slice (\emph{top right}). Both were plotted with a power-law color scale ($\gamma$ = +0.4) to emphasize low density features. \emph{Bottom left}: Confusion matrix between MHD classes (labeled by [sonic, Alfv\'{e}nic] Mach numbers) for the 3D WST-LDA with $J = 6$, $L = 8$, $M = 2$. Matrix entries are the number of test voxels with the corresponding true-predicted label pairing. The grayscale color indicates the percentage of test voxels with the corresponding true label in the test set. In all cases, colors correspond to true labels, as indicated on the vertical axis in the bottom left panel. \emph{Bottom right}: Test and train voxels are plotted in the first two components of LDA space. Circular markers represent test voxels and cross markers represent test voxels that were incorrectly assigned. Slices were taken from an MHD simulation in the ($M_{\rm A}$,$M_{\rm S}$) = (4,2) class.  \label{fig:PPV}}
\end{figure}

The number of coefficients (the dimension of the WST vector) is decreasing as we narrow the range of scales in the above analysis. To test whether the loss of information results explicitly from reduction in the number of coefficients versus loss of information in the image due to binning effectively eliminating certain scales, we generate a series of images of fixed size but with increasing redundancy as follows. We leverage the fact that these simulations have periodic boundary conditions and can therefore repeat a given image in a $2^n \times 2^n$ matrix, downsample the image to $256 \times 256$ resolution, and always compute the WST for $J = 8$, $L = 8$, $M = 2$. These results ($J_{\rm min}$ ``constant'') yield precision scores greater than or equal to $J_{\rm min}$ ``variable'', consistent with the fact that a higher-dimensional representation should contain more information about the image. However, because $J_{\rm min}$ ``constant'' is always within $6\%$ of $J_{\rm min}$ ``variable,'' we conclude that the trend observed in $J_{\rm min}$ ``variable'' is predominantly the result of the scales excluded and not the number of coefficients.

While the highest-resolution maps of the ISM are 2D, the resolution of 3D maps is improving \citep{Green:2019:ApJ:,Leike:2020:A&A:}. Because the physical processes driving the structure of the ISM are happening in 3D, one may expect a measure of non-Gaussianity computed in 3D to capture more information. We show that 3D-WST-LDA ($J = 6$, $L = 8$, $M = 2$) is comparable to 2D-WST-LDA at separating the eight classes in our set of MHD simulations when the same scales are included (Figure \ref{fig:dim}).\footnote{While we only present the $q = 1$ results in Figure \ref{fig:dim} in order to compare to 2D-WST, we find that $q = 1/2$ slightly outperforms $q = 1$ for all $j$ except $j = 7$ while $q = 2$ underperforms compared to $q = 1$ for all $j$. The choice of $q$ (or a set of $q$) used with the WST may be an important optimization for difficult classification problems.} 3D-WST-LDA specifically outperforms 2D-WST-LDA by $\sim10\%$ for $j =$ 5, 6. Note that 3D-WST is a strictly lower dimensional representation for all $j$ studied as we do not compute second-order coefficients between different angular bins in the 3D-WST procedure.\footnote{We note that the multipole expansion with $L_{\rm max}$ used for the 3PCF has one more angular component as compared to 2D-WST with the same $L$. Excluding the zero frequency component, the multipole expansion with $L_{\rm max}$ contains angular frequencies $2\pi/n$ for $n$ an integer from $1$ to $L_{\rm max}$. For 2D-WST the number of angular divisions between $-\pi/2$ and $\pi/2$ sets the Nyquist frequency to be $\pi/L$. Since the angular scales for 3PCF and 2D-WST are not identical, we simply compare the number of coefficients.} 2D-WST and 3D-WST also use different wavelets (see Equation \ref{eq:WST_2D_wavelet} and \ref{eq:WST_3D_wavelet}) and so we leave a more detailed comparison of the different bases to future work. 3D-WST-LDA clearly can provide similar classifications to 2D-WST-LDA, though the increased computational expense means it will likely only be used when 3D correlations are necessary (see Appendix \ref{sec:CompCost} for computational costs).

\subsection{Robustness of Classification} \label{sec:Robust}
\subsubsection{Using Velocity Information}
While few 3D maps of the ISM exist in position-position-position (PPP) space and are of relatively low resolution (1 pc; \citealt{Leike:2020:A&A:}), high resolution maps of gas tracers (\HI, ${}^{12}$CO, ${}^{13}$CO) in position-position-radial velocity (PPV) space are available \citep{Dame:2001:ApJ:,Ridge:2006:AJ:,Pineda:2008:ApJ:,Peek:2018:ApJS:,Clark:2019:ApJ:}. These PPV maps are also used to constrain sonic Mach numbers of molecular and atomic clouds \citep{Heiles:2003:ApJS:,Burkhart:2010:ApJ:,Kainulainen:2013:A&A:}. We convert our MHD simulations to PPV space by interpolation using the density field and $x$-velocity field.\footnote{Similar results were obtained for $y$- and $z$-velocity field PPV cubes.} Representative 2D slices in PP and PV space are shown in Figure \ref{fig:PPV}. Classification on 2D slices was difficult (average precision $45\%$), but by utilizing the 3D WST ($J = 6$, $L = 8$, $M = 2$) we achieved a precision of 76\%.\footnote{$J = 7$ for the 3D WST had too few samples for a reliable test-train split.} We suspect the reason 2D classification was difficult is the little correlation between density and velocity fields observed by \citet{Burkhart:2009:ApJ:}. However, those authors noted a correlation between Mach numbers and velocity dispersion, which may explain why the 3D WST in PPV space, which captures a large range of velocities, performs well. This ability of WST-LDA to operate in PPV space enables application both to a wider variety of data and to higher resolution 3D data. This higher resolution PPV data may be imperative for mapping regions with overlapping clouds observed in dust and gas tracers. Given that a large range of velocities was needed to determine the Mach number, resolving overlapping clouds may be difficult unless the clouds are at well-separated in velocity space. 

\newpage
\subsubsection{Artifacts and Missing Data}
Pattern noise is a common feature in imaging data sets. Long-term sensitivity drift in scanning detectors produced stripe artifacts in IRAS \citep{Wheelock:1994:STIN:} and \textsc{AKARI} \citep{Doi:2015:PASJ:}. A ground loop in the readout electronics of a charge-coupled device (CCD) can imprint nearby radio transmissions on the data.\footnote{One of the authors detected a Maui radio station with a Panoramic Survey Telescope and Rapid Response System (Pan-STARRS) CCD.} Even highly sophisticated instruments on the Hubble Space Telescope can have electronics problems that leave subtle ``herringbone'' patterns in the data \citep{Jansen:2010:hstc:}. A desideratum of a classifier is that it be robust to such pattern noise, or at least be able to warn that it is present. 

\begin{figure}[ht!]
\includegraphics[width=\linewidth]{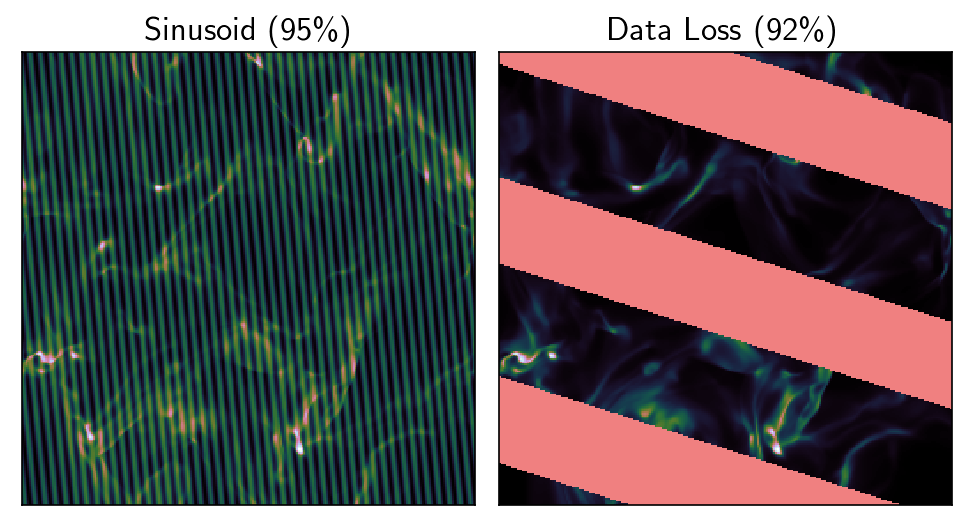}
\caption{Example of additive sinusoidal noise (\emph{left}) and our data elimination scheme (\emph{right}). Removed data was set to light red on our color scale for illustration purposes, but was set to zero when computing WST-LDA. The precision score for the MHD classification task is noted in the title. Representative slices were taken from an MHD simulation in the ($M_{\rm A}$, $M_{\rm S}$) = (4, 2) class. In both cases quantitative classification is evidently robust to these artifacts, because the precision is comparable to the noise-free value of 94\%.  \label{fig:artifact}}
\end{figure}

For testing purposes, we mimic pattern noise by adding a sinusoidal pattern with random wavevector to the MHD slices. Specifically, we generate for each MHD slice (6,912 draws) a random amplitude (10-30\% of signal max), phase, period width (2-25 pixels), and angle relative to the $x$ and $y$ axis and add it to the MHD slice prior to computing the WST coefficients (Figure \ref{fig:artifact}). In some cases, large contiguous regions of an image may be lost due to satellite transits, diffraction spikes, or ghosts of bright stars. We model this by introducing a cutoff for eliminating pixels based on the sinusoidal field used above (except with periods in the range of 43-256 pixels and amplitudes 10-20\%). The sinusoidal field was not added to the data, but the values in the image were set to zero for all indices where the sinusoidal field exceeded 10\% (Figure \ref{fig:artifact}). In both cases we observe at most a slight reduction in the precision (95\% and 92\% respectively, compared to 94\%) and increased variance for each class. However, the quantitative classification is evidently robust to these artifacts, and the qualitative structure of the clusters in the space of the first three LDA components was also unaffected.

\subsubsection{Dependence on Point-Spread Function}
In observational images, the true signal is also convolved with a point spread function (PSF). We model this as a Gaussian PSF with a FWHM of $2.355\sigma$ (in pixels). Further, we do not assume the PSF is rotationally symmetric and allow for nonzero aspect ratios AR $= \sigma_x/\sigma_y$. We investigated the following PSFs in detail: ($\sigma$, AR) = (2,1), (2,2), (4,1), (4,2), (4,3). The MHD classification by WST-LDA for samples from a single PSF was qualitatively unchanged and had a precision $>92\%$. Further, when all the WST coefficients from the PSF-blurred images and original images were combined, LDA was able to classify the images on Mach numbers with a precision of 96\%. 

\begin{figure}[hb!]
\includegraphics[width=\linewidth]{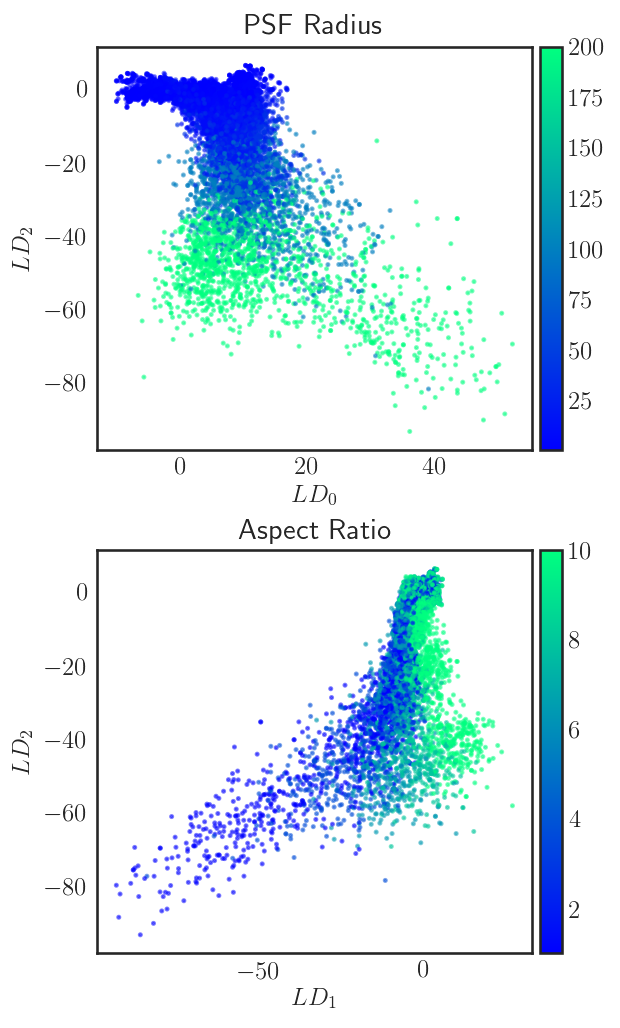}
\caption{Effect of PSF radius and aspect ratio in a representative WST-LDA space for classification of MHD simulations. Each point comes from applying WST to a 2D image from our MHD simulations convolved with a PSF of the indicated radius (\emph{top}) and aspect ratio (\emph{bottom}) given by the color bar. An interactive version of this figure is available at \url{https://faun.rc.fas.harvard.edu/saydjari/RWST_2020/MHD_PSF.html} showing the PSF and aspect ratio dependence of the position of each point in the first three LDA coordinates. The MHD classification clusters for images with PSF with parameters (2,1) are also shown in the LDA basis for the unblurred images and vice versa. \label{fig:PSF}}
\end{figure}

However, we must emphasize that the optimal LDA space for each individual PSF is different. For instance, the LDA space for (2,1) and the LDA space for the original images are not the same. In the interactive version of Figure \ref{fig:PSF}, we demonstrate how the WST coefficients for (2,1) look in the LDA space for the original images and vice versa. While these spaces are not identical, the helix that we observed in the native LDA space is still present, but shifted and compressed or expanded. We observe that for small $\sigma$ and AR, the old LDA space can be approximately continuously deformed into the new LDA space (not shown). 

Ideally, we would find a classification space which is as insensitive as possible to the PSF. To approximate this, we take the original images in addition to five blurred copies ((2,1), (2,2), (4,1), (4,2), (4,3)) and define the WST-LDA space for the Mach number classification task on that entire set. To better understand how the position of an image's WST coefficients in the LDA space depend on the $\sigma$ and AR of the PSFs, we take four random slices from each timestep of each MHD class and compute the WST coefficients for a Cartesian product of $\sigma$ and AR. We then plot the WST coefficients corresponding to each each image with a given PSF in the space defined above in Figure \ref{fig:PSF}.\footnote{We do not scale the WST coefficients to have mean zero and unit variance prior to computing the LDA space so we can more easily study which coefficients are most affected by the PSF.} We observe that increasing the PSF radius moves images to lower $LD_2$ compared to the original helical cluster. On this tail induced by the large PSF radii, the images are separated along $LD_1$ by AR. Thus both $\sigma$ and AR are encoded in the WST-LDA space. Similarly, the WST-LDA spaces for $J_{\rm min}$ = 0 and $J_{\rm min}$ = 1 are related, but not identical, suggesting that the scale of a feature relative to the minimum pixel scale will also deform the WST-LDA space. We foresee creating a classifier which is invariant to, or deforms in a known way in response to, changes in the PSF and pixel scale. This will likely be the next step before WST-LDA can be applied to classify relevant hydrodynamic parameters (or any other parameter of interest) on observational data.

\begin{figure}[ht!]
\includegraphics[width=\linewidth]{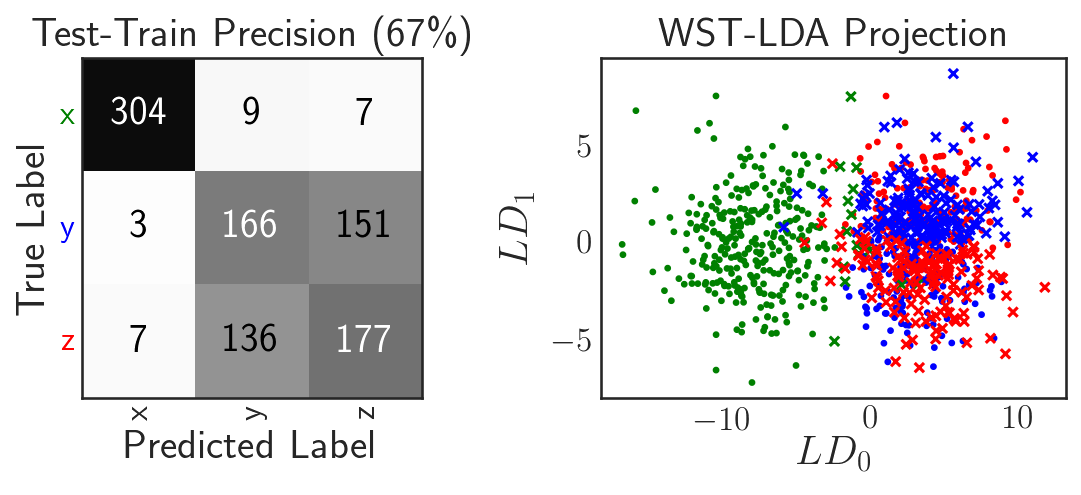}
\caption{Confusion matrix between slices taken from the $x$, $y$, and $z$ axes for the sub-Alfv\'{e}nic MHD classes ($M_{\rm A} = 0.7$) (\emph{left}). Matrix entries are the number of test images with the corresponding true-predicted label pairing. The grayscale color indicates the percentage of test images with the corresponding true label in the test set. In all cases, colors correspond to true labels, as indicated on the vertical axis in the left panel. Test and train images are plotted in the first two components of LDA space (\emph{right}). Circular markers represent test images and cross markers represent test images that were incorrectly assigned. WST-LDA can clearly distinguish the magnetic field axis, in the absence of magnetic field wandering, from the density field alone. \label{fig:axesSep}}
\end{figure}

\subsection{Sensitivity to Magnetic Field Direction} \label{sec:Sense}

In the MHD turbulence simulations, the original magnetic field direction in the simulation box is topologically constrained by the periodic boundary conditions of the simulation. This introduces a preferred direction in the simulation because the characteristics and scaling of turbulence parallel and perpendicular to the magnetic field depend on the Alfv\'{e}nic state of the gas \citep{Goldreich:1995:ApJ:}. The mean magnetic field remains relatively unperturbed in the sub-Alfv\'{e}nic regime, in contrast to the super-Alfv\'{e}nic regime where the direction of field lines can wander \citep{Lazarian:1999:ApJ:, Cho:2002:ApJ:}. In the simulations studied here, the mean magnetic field is initialized along the $x$-axis. We thus pooled for all sub-Alfv\'{e}nic MHD classes ($M_{\rm A} = 0.7$) density slices taken perpendicular to the $x$, $y$, and $z$ axes and show that WST-LDA picks out the $x$-axis as distinct from the $y$- and $z$-axes (Figure \ref{fig:axesSep}). While the precision for separating $x$, $y$, and $z$ is shown in the Figure, the fact that we find a precision of $98\%$ for differentiating $y$ and $z$ from $x$ is a more meaningful conclusion in light of the degeneracy of $y$ and $z$. For the super-Alfv\'{e}nic MHD classes ($M_{\rm A} = 2.0$), WST-LDA returns a precision of only $54\%$. This reduced precision compared to the sub-Alfv\'{e}nic classes indicates that WST-LDA is sensitive to the effect of magnetic field wandering; the signatures of the magnetic field present in the density field for the sub-Alfv\'{e}nic classes were no longer present. Thus, WST-LDA may provide a insight into the direction of homogeneous magnetic fields in the ISM.

\newpage
\section{Conclusions} \label{sec:Conc}

We introduce WST-LDA as a new low-dimensional classifier for non-Gaussianity. In comparing to the RWST and 3PCF, we find the WST and RWST more easily differentiate non-Gaussian from Gaussian fields and classify MHD simulation parameters. RWST-LDA appears more robust than WST-LDA to sums along the line of sight and achieves higher precision, but is susceptible to outliers in its current form. To our knowledge this is the first classification of MHD Mach numbers with $\sim$97\% precision achieved without training a neural network. We show that the power of WST-LDA comes mostly from the number (and the relevance) of scales included in the scattering transform, not the number of coefficients. We similarly demonstrate 3D-WST-LDA and its application to PPV space, a case where the 3D correlations are essential to high-precision classification. Classification in PPV space may be improved by development of anisotropic 3D wavelets preferring the velocity axis. The classification by WST-LDA was robust to additive sinusoidal noise, partial data loss, and convolution with various point spread functions. However, the WST-LDA space in which the classification occurs was not invariant to the aforementioned effects or the scale of features relative to the minimum pixel scale. Thus, modifying WST-LDA to be sensitive only to the MHD parameters and not details of the resolution and pixel scale of measurements will be the next step before classification of observational data. 

\section{Code and Data Availability} \label{sec:dataavil}

Data products associated with the paper are publicly available at \doi{10.5281/zenodo.4057157}. This includes the computed WST, RWST, and 3PCF coefficients, our selected subset of SFD, PPV density cubes from the MHD simulations, and our code for computing the 2D 3PCF. Reuse of our 3PCF code should cite \citealt{Slepian:2015:MNRAS:,Slepian:2016:MNRAS:,Portillo:2018:ApJ:,3PCFkeyunpub}. \textsc{Jupyter} notebooks containing code to reproduce all figures in the text, some minimal working examples, and how to run these computations on a cluster are also included. The full MHD simulation runs are available from the Catalog for Astrophysical Turbulence Simulations (CATS) at \url{http://www.mhdturbulence.com}.

\newpage
\acknowledgments
A.S. gratefully acknowledges support by a National Science Foundation Graduate Research Fellowship (DGE-1745303). D.F. acknowledges support by NSF grant AST-1614941, “Exploring the Galaxy: 3-Dimensional Structure and Stellar Streams.” B.B. is grateful for support from the Simons Foundation. S.K.N.P. acknowledges support from the DIRAC Institute in the Department of Astronomy at the University of Washington. The DIRAC Institute is supported through generous gifts from the Charles and Lisa Simonyi Fund for Arts and Sciences, and the Washington Research Foundation. We acknowledge Catherine Zucker and Michael Foley for helpful discussions. A.S. acknowledges Sophia S\'{a}nchez-Maes for helpful discussion and much support. Computations in this paper were run on the FASRC Cannon cluster supported by the FAS Division of Science Research Computing Group at Harvard University.

\facilities{IRAS, COBE}

\software{
\textsc{astropy} \citep{AstropyCollaboration:2013:A&A:},
\textsc{dustmaps} \citep{Green:2018:JOSS:}, 
\textsc{glue} \citep{Beaumont:2015:ASPC:,Robitaille:2017:zndo:}
\textsc{h5py} \citep{collette_python_hdf5_2014}, 
\textsc{ipython} \citep{Perez:2007:CSE:}, 
\textsc{kymatio} \citep{Andreux:2018:arXiv:}, 
\textsc{matplotlib} \citep{Hunter:2007:CSE:}, 
\textsc{numpy} \citep{vanderWalt:2011:CSE:}, 
\textsc{pytorch} \citep{Paszke:2019:arXiv:}, 
\textsc{scipy} \citep{Virtanen:2020:NatMe:}, 
\textsc{scikit-learn} \citep{Pedregosa:2012:arXiv:}, 
\textsc{scikit-image} \citep{vanderWalt:2014:arXiv:}, 
\textsc{yt} \citep{Turk:2011:ApJS:}, 
}
\onecolumngrid

\appendix
\section{Computational Costs} \label{sec:CompCost}
\begin{deluxetable}{c|DDD|DDD}[ht!]
\tablenum{1}
\tablecaption{Computational cost for the 2D-WST, 3D-WST, and 3PCF \label{tab:compCost}}
\tablewidth{0pt}
\tablehead{
\multirow{2}{*}{\textbf{J}} & \multicolumn{6}{c|}{Time/Coeff (core-milliseconds)} & \multicolumn{6}{c}{Number of Coefficients} \\
& \multicolumn2c{2D-WST}& \multicolumn2c{3D-WST} & \multicolumn2c{3PCF} \vline & \multicolumn2c{2D-WST}& \multicolumn2c{3D-WST} & \multicolumn2c{3PCF}
}
\decimals
\startdata
\textbf{8} & $3.4$ & $2600$ & $1.2$ & 1857 & 1215 & 360 \\
\textbf{7} & $1.4$ & $140$ & $0.95$ & 1401 & 972 & 280 \\
\textbf{6} & $0.75$ & $12$ & $0.73$ & 1009 & 756 & 210 \\
\textbf{5} & $0.47$ & $1.5$ & $0.56$ & 681 & 567 & 150 \\
\textbf{4} & $0.30$ & $0.35$ & $0.39$ & 417 & 405 & 100 \\
\textbf{3} & $0.23$ & $0.12$ & $0.27$ & 217 & 270 & 60 \\
\enddata
\end{deluxetable}

To provide an idea of relative computational cost for each of the non-Gaussian descriptors used, we present the number of coefficients and computational time per coefficient for the 2D-WST, 3D-WST, and 3PCF (Table \ref{tab:compCost}). The computation of the RWST coefficients from the WST coefficients is just the usual cost of least-squares optimization (and relatively fast compared to the WST computation). Recall that the definition of $J$ for each method is slightly different. The 2D-WST computed up to a scale J includes $j = 0, ..., J-1$ while the 3D-WST and 3PCF include $j = 0, ..., J$. For the 2D and 3D-WST we used $L = 8$ and for the 3PCF we used $L = 9$. For $J \geq 7$, the memory load for 3D-WST exceeds 100 MiBs and needs to be considered. For 3D-WST and 3PCF, the coefficients were calculated on an image of uniform grid size $2^J$. For 2D-WST only, the computational time reflects calculating the coefficients on $2^{8-J} \times 2^{8-J}$ subsets of the full $256^2$ images of size $2^{J} \times 2^{J}$. These experiments were executed on the FASRC Cannon cluster at Harvard University on a compute node with water-cooled Intel 24-core Platinum 8268 Cascade Lake CPUs with 192GB RAM running 64-bit CentOS 7. Cascade Lake cores have dual AVX-512 fused multiply-add (FMA) units. Code scaling is reported on a single core in a \textsc{Python} environment specified by the \textsc{yaml} file in Section \ref{sec:dataavil}.

\section{Effect of Mean Zero on 3PCF} \label{sec:3PCF2PCF}
Defining a density fluctuation field with mean zero as $\delta(\vec{x}) \equiv \rho(\vec{x})/\bar{\rho} -1$, with $\bar{\rho}$ the average density, the 3PCF is 
\begin{ceqn}
\begin{align}
\left< \delta(\vec{x}) \delta(\vec{x}_1 + \vec{r}_1) \delta(\vec{x}+ \vec{r}_2)\right>
\end{align}
\end{ceqn}
where the angle brackets denote averaging over translations and rotations. Substituting in the definition of $\delta$ and defining the $density$ 2PCF as $\xi_{\rho}(s) \equiv \left< \rho(\vec{x})\rho(\vec{x} + \vec{s})\right>$, we find 
\begin{ceqn}
\begin{align}
\left< \delta(\vec{x}) \delta(\vec{x}+\vec{r}_1) \delta(\vec{x} + \vec{r}_2)\right> =\zeta_{\rho}(r_1, r_2,\hat{r}_1 \cdot \hat{r}_2)  - \left[\xi_{\rho}(r_1) + \xi_{\rho}(r_2) + \xi_{\rho}(r_3)\right]
\end{align}
\end{ceqn}
where $\zeta_{\rho}$ is the 3PCF of the $density$ field and we used that $|\vec{r}_2 - \vec{r}_1| = r_3$. So as claimed, taking a zero-mean field results in subtracting off the 2PCF contribution to the 3PCF.

\section{RWST Angle Dependence} \label{sec:GRFosc}

\begin{figure}[ht!] \centering
\includegraphics[width=0.9\textwidth,height=0.9\textheight,keepaspectratio]{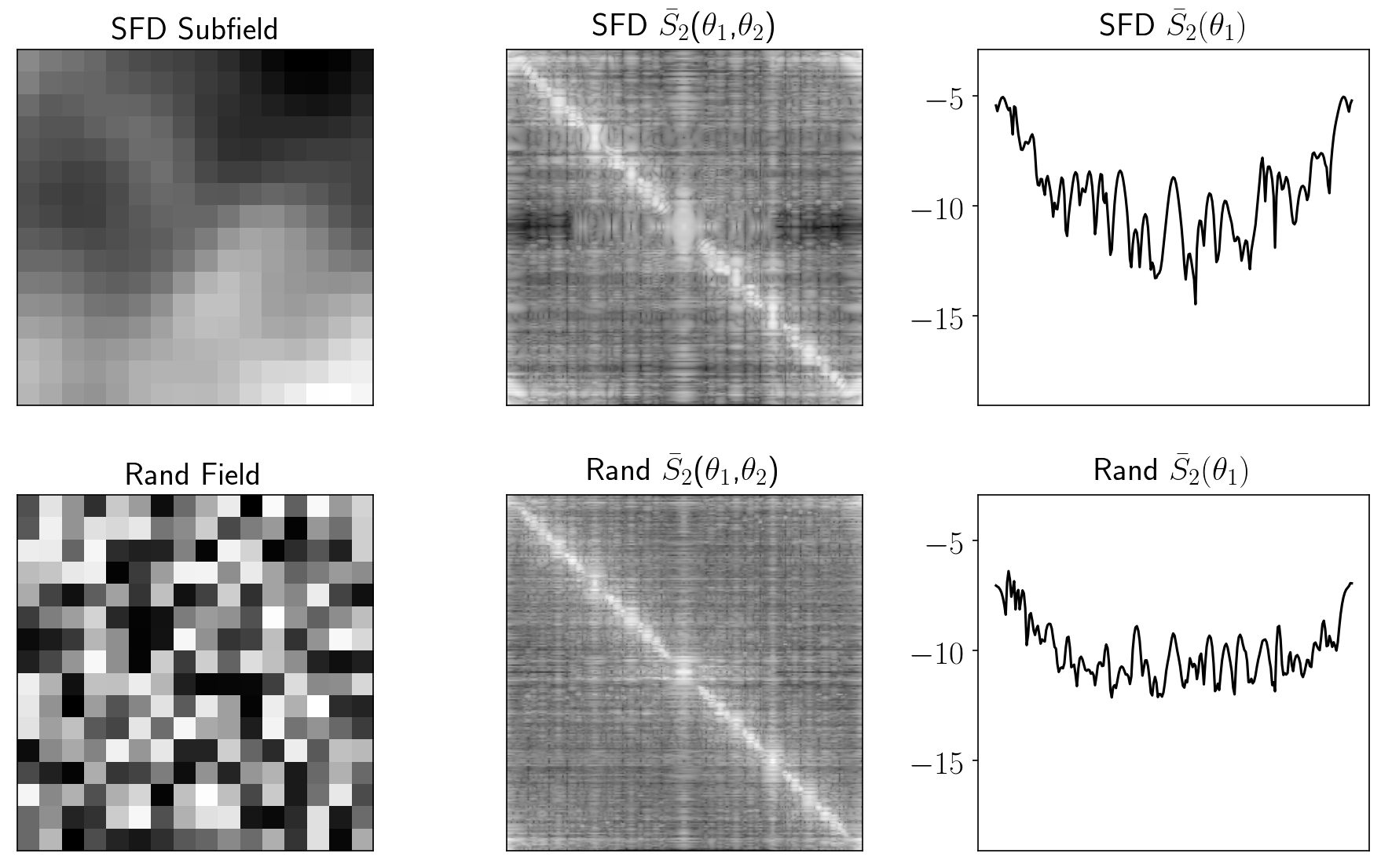}
\caption{Images used for computing fine angle dependence on an SFD subfield (\emph{top left}) and random pixel image (\emph{bottom left}). The second order WST coefficient map in terms of $\theta_1$ and $\theta_2$ is shown for SFD (\emph{top middle}) and Rand (\emph{bottom middle}). On the right, line cuts along $\theta_1$ are plotted for SFD (\emph{top right}) and Rand (\emph{bottom right}) with the same y-scale spanning the minimum and maximum $\bar S_2$ obtained for both images. \label{fig:GRFoscFig}}
\end{figure}

While we show that the RWST transformation of the WST coefficients can improve the performance of RWST-LDA compared to WST-LDA, the origin of this improvement remains not fully understood. To illustrate this, we compute the WST coefficients $J = 2, L = 256, M = 2$ for two images that are $16 \times 16$ pixels: a subset of an SFD image and an image with random pixels. The SFD image at this pixel scale provides only an example of a real, slowly varying field in contrast to the random image in which neighboring pixels are totally uncorrelated. The angle dependence of the second order WST coefficients for an internal $4 \times 4$ subset of pixels is shown as a function of $\theta_1$ and $\theta_2$ in Figure \ref{fig:GRFoscFig}. A horizontal line cut near the top of the WST coefficient map is also shown to illustrate the oscillation envelope fit by the cosine in the RWST model. While detailed aspects of the WST coefficient maps are different, both the continuously varying field and the random image display clear oscillations. It is possible much of the relevant information enters into the amplitude of the WST coefficients, but more work is needed to understand the optimal parameterization of these oscillations and better understand their interpretation.

\section{Classification of Cumulative Sums} \label{sec:cumSum}

In order to mimic 2D dust maps which measure dust integrated along the lines of sight, we sampled the cumulative sum of our MHD density cubes along each axis and repeated the classification based on sonic and Alfv\'{e}nic Mach numbers using LDA combined with the WST, RWST, and 3PCF. WST-LDA achieves a precision of only 54\% and shows a large broadening in the variance for each class, which leads to significant cross-talk between multiple adjacent classes. However, the confusion matrix is still predominately diagonal with most of the density within two classes from the correct label. Thus, sampling the cumulative sum can be viewed as causing an increased uncertainty in the Mach number assignments. RWST-LDA shows a much smaller decrease in precision (to 81\%) and much smaller class variance on the cumulative sum images compared to WST-LDA. This further supports the claim that the nonlinear transformation of the WST coefficients made by the RWST is not only useful for dimension reduction, but also makes the classifier more robust against variations uninformative to distinguishing the MHD classes. While the 3PCF-LDA precision drops to approximately to 51\%, near that for WST-LDA, the classification is much worse in that many images overlap in the LDA space and are all predicted to be in the same class. This is a function of the nonlinear spacing of classes in the 3PCF-LDA space. The comparatively poor performance of 3PCF also suggests that the specific combination of up to $4^{th}$ order moments included in the WST is well chosen since the 3PCF alone is so strongly modified.

\begin{figure*}[!hb] \centering
\includegraphics[width=0.65\textwidth,height=0.65\textheight,keepaspectratio]{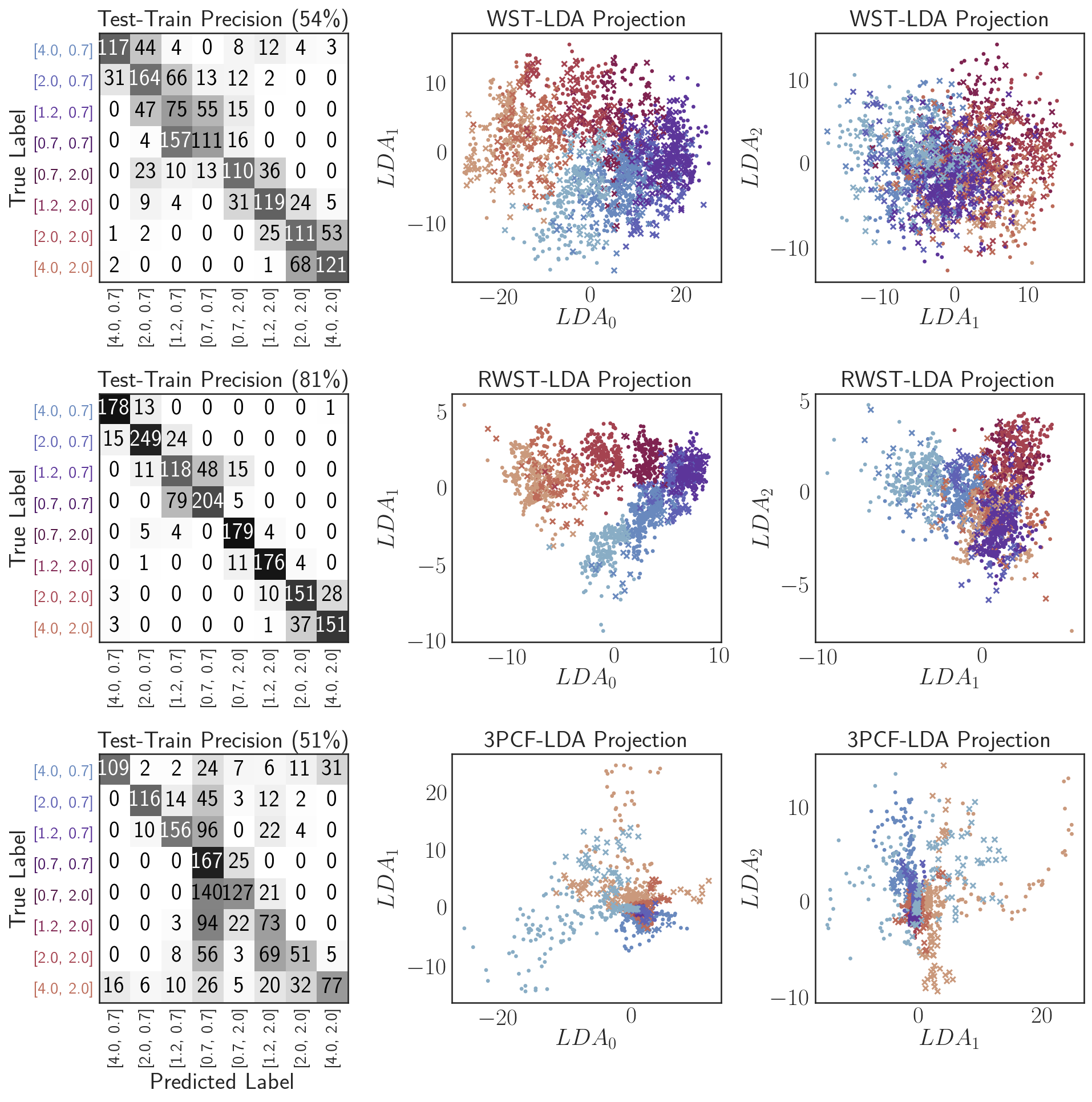}
\caption{
Confusion matrix between MHD classes labeled by [sonic, Alfv\'{e}nic] Mach numbers when applying WST-LDA (\emph{top left}), RWST-LDA (\emph{middle left}), and 3PCF-LDA (\emph{bottom left}) for images drawn from cumulative sums along the MHD simulation density cubes. Matrix entries are the number of test images with the corresponding true-predicted label pairing. The grayscale color indicates the true postitive percentage relative to the number of test images with the corresponding true label. Test images are shown in the first two components of LDA space for the WST (\emph{top right}), RWST (\emph{middle right}), and 3PCF (\emph{bottom right}). Circle markers represent images that were correctly assigned, and cross markers represent those incorrectly assigned. In all cases, colors correspond to true labels. An interactive version of this figure is available at \url{https://faun.rc.fas.harvard.edu/saydjari/RWST_2020/MHD_LDA_cumsum.html} showing the images in 3D LDA space and allowing isolation of the train, correctly labeled test images, and incorrectly labeled test images. \label{fig:WST_LDA_MHD_cumsum}}
\end{figure*}

\pagebreak

\bibliography{WST.bib}{}
\bibliographystyle{aasjournal}
\end{document}